\documentclass[pdflatex,sn-chicago]{sn-jnl}

\usepackage{lmodern}
\usepackage{array,multirow,diagbox}
\newcolumntype{P}[1]{>{\centering\arraybackslash}p{#1}}
\usepackage{amsmath,amssymb,mathtools,bm}
\usepackage{siunitx}
\usepackage{graphicx}
\usepackage{overpic}
\usepackage{pifont}
\usepackage{hyperref}
\usepackage{cleveref}

\newcommand{\xmark}{\ding{55}}
\newcommand{\aref}[1]{\hyperref[#1]{Appendix}}
\newcommand{\diff}[2]{\frac{\mathrm{d} #1}{\mathrm{d} #2}}
\newcommand{\pdiff}[2]{\frac{\partial #1}{\partial #2}}

\newcommand{\intd}{\mathop{}\!\mathrm{d}}
\newcommand{\lap}[1]{\nabla^2{#1}}
\newcommand{\tensor}[1]{\bm{#1}}
\newcommand{\diagMat}[3]{\begin{bmatrix}
        #1 & 0 & 0\\
        0 & #2 & 0\\
        0 & 0 & #3
    \end{bmatrix}}
\renewcommand{\div}[1]{\nabla\cdot{#1}}
\newcommand{\at}[2]{\left.{#1}\right\rvert_{#2}}
\newcommand{\eye}{\tensor{I}}

\DeclareSIUnit{\day}{day}

\newcommand{\tableheader}{\multirow{2}{*}{Model} & Mechanical feedback on growth & Guaranteed steady state of radius & Canonical growth curves & Bounded solid stress & Plausible residual stress profiles}

\newcommand{\initialOuterRadius}{B}
\newcommand{\outerRadius}{b}
\newcommand{\elasticStretch}{\alpha}
\newcommand{\growthStretch}{\gamma}
\newcommand{\shearModulus}{\mu}
\newcommand{\springConstant}{\kappa}
\newcommand{\growthRateConst}{k}
\newcommand{\stressModifier}{n}

\newcommand{\necrosisThreshold}{\hat{c}}
\newcommand{\diffusionCoeff}{D}
\newcommand{\consumptionRate}{\lambda}
\newcommand{\boundaryNutrient}{c_{\infty}}
\newcommand{\diffusiveLengthscale}{L}
\newcommand{\outerRadiusThreshold}{\hat{\outerRadius}}
\newcommand{\cI}{c_{\mathrm{I}}}
\newcommand{\cII}{c_{\mathrm{II}}}
\newcommand{\rThreshold}{\hat{r}}

\newcommand{\stressThreshold}{\hat{\sigma}}

\newcommand{\dummy}[1]{\tilde{#1}}
\newcommand{\ststNutrient}{\outerRadius_{N}^{*}}

\newcommand{\gobble}[1]{}

\newcommand{\changed}[1]{#1}

\jyear{2022}

\begin{document}

\title[Minimal Morphoelastic Models of Solid Tumour Spheroids: A Tutorial]{\changed{Minimal Morphoelastic Models of Solid Tumour Spheroids: A Tutorial}}

\author*[1]{\fnm{Benjamin J.} \sur{Walker}}\email{benjamin.walker@ucl.ac.uk}
\author[1,2]{\fnm{Giulia L.} \sur{Celora}}\email{g.celora@ucl.ac.uk}
\author[2]{\fnm{Alain} \sur{Goriely}}\email{goriely@maths.ox.ac.uk}
\author[2]{\fnm{Derek E.} \sur{Moulton}}\email{moulton@maths.ox.ac.uk}
\author[3]{\fnm{Helen M.} \sur{Byrne}}\email{helen.byrne@maths.ox.ac.uk}

\affil*[1]{\orgdiv{Department of Mathematics}, \orgname{University College London}, \orgaddress{\street{Gordon Street}, \city{London}, \postcode{WC1H 0AY}, \country{United Kingdom}}}
\affil[2]{\orgdiv{Wolfson Centre for Mathematical Biology, Mathematical Institute}, \orgname{University of Oxford}, \orgaddress{\street{Woodstock Road}, \city{Oxford}, \postcode{OX2 6GG}, \country{United Kingdom}}}

\abstract{\changed{Tumour spheroids have been the focus of a variety of mathematical models, ranging from Greenspan's classical study of the 1970s through to contemporary agent-based models. Of the many factors that regulate spheroid growth, mechanical effects are perhaps some of the least studied, both theoretically and experimentally, though experimental enquiry has established their significance to tumour growth dynamics. In this tutorial, we formulate a hierarchy of mathematical models of increasing complexity to explore the role of mechanics in spheroid growth, all the while seeking to retain desirable simplicity and analytical tractability. Beginning with the theory of morphoelasticity, which combines solid mechanics and growth, we successively refine our assumptions to develop a somewhat minimal model of mechanically regulated spheroid growth that is free from many unphysical and undesirable behaviours. In doing so, we will see how iterating upon simple models can provide rigorous guarantees of emergent behaviour, which are often precluded by existing, more complex modelling approaches. Perhaps surprisingly, we also demonstrate that the final model considered in this tutorial agrees favourably with classical experimental results, highlighting the potential for simple models to provide mechanistic insight whilst also serving as mathematical examples.}}


\maketitle

\section{Introduction}\label{sec: intro}
Cancer is a disease that impacts the lives of tens of millions of people
worldwide each year and represents a leading cause of death \citep
{Sung2021}. The growing prevalence and severity of the disease has driven
rapid advancements in our understanding of the biology that underpins tumour
growth, as highlighted by the evolving characterisation of the Hallmarks of
Cancer in the renowned works of \citet
{Hanahan2000,Hanahan2011}. A subset of this vast body of research has
considered the broad range of stimuli that are known to affect the behaviour
of biological cells and tissues, including cancer cells. These stimuli
include, but are not limited to, the availability of nutrients for growth,
mechanical forces acting on tissues, and electric fields \citep
{Vaupel1989,Pavlova2016,Northcott2018,Sengupta2018,Kolosnjaj-Tabi2019}.
Whilst the study of a single stimulus is often difficult or intractable in
vivo, experimental assays have provided a means to focus on one or two
stimuli at a time. For instance, a common approach for studying the early
stages of avascular tumour growth is to consider three-dimensional
collections of cancer cells known as tumour spheroids \citep
{Hirschhaeuser2010}, which are thought to better emulate in vivo environments
than alternative two-dimensional assays whilst still enabling the targeted
study of tumour growth stimuli. Spheroid assays have been used to study the
impact on tumour growth of multiple stimuli, such as how nutrient
availability affects cancer development \citep
{Kunz-Schughart1998,Murphy2022}, and to gain insight into the mechanical
inhibition of growth, as in the now-classical work of \citet
{Helmlinger1997}, which we consider in detail below.

In addition to the range of experimental investigations that have involved the
use of tumour spheroids, a multitude of mathematical models have been
developed to study spheroids. These efforts, which are part of the emergent
field of  \emph{mathematical oncology}, range from simple single-compartment
ordinary differential equation (ODE) models to complex, multiscale schemes
and hybridised partial differential equation (PDE) and agent-based methods,
which differ in complexity, spatial resolution, and scale. One of the
earliest and best known models is that of \citet
{Greenspan1972}, which considers how the composition of a tumour spheroid
evolves as the growing tumour limits the availability of diffusing nutrients
(in this case oxygen) to the central core of the spheroid. Greenspan's PDE
approach, in which the behaviour of cells is driven by the local nutrient
concentration, has since been adapted by many authors and adds to the breadth
of mathematical methods that have been employed in the study of cancer. The
reviews of \citet{Araujo2004}, \citet
{Roose2007}, and \citet{Bull2022} provide a comprehensive
summary of these theoretical approaches.

Since Greenspan's early work, many mathematical models have focussed on
exploring how nutrient availability and spatial constraints limit tumour
growth \citep{Ward1997,Sherratt2001,Murphy2022}. In contrast, however, the
notion of mechanical feedback remains relatively unexplored in theoretical
works, despite mechanical effects being increasingly appreciated as
significant in many biological settings. For instance, the work of \citet{Helmlinger1997} demonstrated that mechanical resistance to
growth can markedly limit the growth of tumour spheroids, with resistance in
this case being imparted via an agarose gel that surrounds the spheroids.
More recent experimental studies add weight to \citeauthor{Helmlinger1997}'s
conclusions, such as that of \citet{Cheng2009}, which considered
the effects of externally imposed stresses on tumours and measured the
impacts of mechanical stress on cell proliferation and apoptosis.
Notwithstanding these experimental results, there is no consensus about \emph
{how} mechanical cues alter growth dynamics on the tissue and cell scales.
This uncertainty has spawned a range of phenomenological continuum models of
mechanically influenced tumour growth \citep
{Chen2001,Roose2003,Byrne2009,Ambrosi2004,Ambrosi2002,Ambrosi2017,Ambrosi2009,Byrne2003},
which have successfully reproduced both tumour growth curves and profiles of
accumulated solid stress \citep{Nia2018}. These theoretical studies have made
different modelling choices, most notably in the posited constitutive
couplings between mechanics and growth. A key challenge is finding a coupling
that represents the least complex relation needed to generate experimentally
observed profiles of growth and stress. Such simplicity is often desirable in
mathematical models when detailed understanding of the biological mechanisms
is lacking, and such `minimal ingredients' models generally facilitate both
ready interpretation and analytical study, the latter of which can provide
rigorous characterisation of model dynamics and behaviours. Such
characterisations are largely absent from existing solid-mechanical models of
spheroid growth. In particular, it remains to be established whether
agreement between numerical solutions of existing mechanical tumour models
and experimental data depends strongly on the particular parameter regimes
employed, or whether they reproduce the observed phenomena more generally.

Motivated by these observations, the \changed{scientific} aim of this \changed
{tutorial} is to develop a minimal model of tumour spheroid growth that
reproduces observed growth dynamics, under varying external conditions, and
permits rigorous characterisation of model behaviours. \changed{In pursuit of
this goal, we will adopt an iterative and expository approach to model
development, beginning with a simple, established foundation and successively
posing extensions and modifications in order to realise a number of desirable
properties. To facilitate the development of such a simple mathematical
model, each of our models will be based on} the solid-mechanical framework
of \emph{morphoelasticity}, introduced by \citet
{Rodriguez1994} and reviewed by \citet{Ambrosi2011} and
\citet{Kuhl2014}. The theory of morphoelasticity has been applied broadly
to problems of biological growth and often leads to models that are
analytically tractable and numerically straightforward, such as a recent
model of the human eye \citep{Kimpton2021}, as described by \citet
{Goriely2017}.

\begin{figure}[t]
    \centering
    \includegraphics[width=0.6\textwidth]{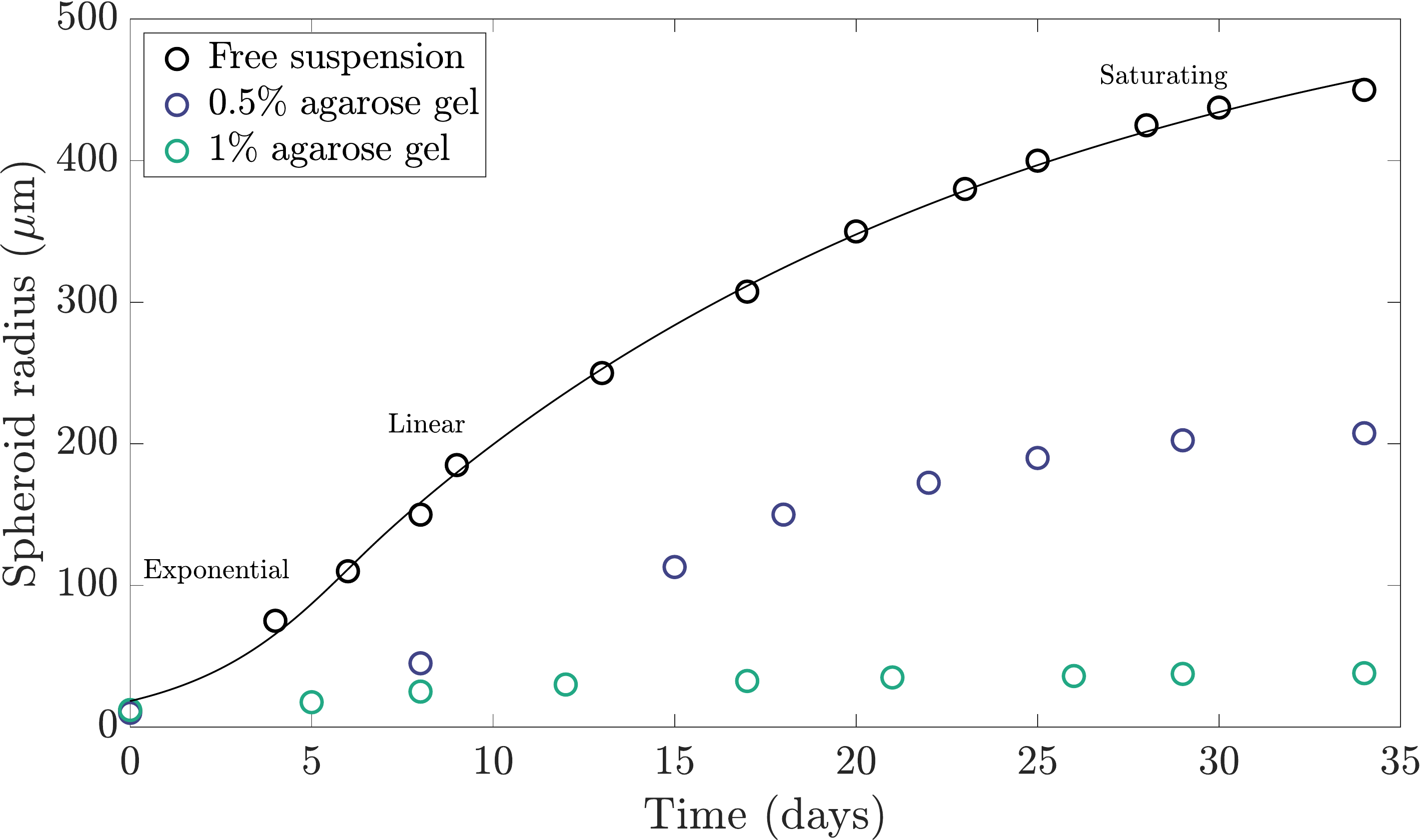}
    \caption{Mechanically influenced tumour spheroid growth. Experimentally observed growth dynamics of tumour spheroids in various extracellular media are indicated by empty circles, as reported by \citet{Helmlinger1997}, digitised by \citet{Yan2021}. Tumours growing in agarose gels of higher concentration experience reduced growth compared to evolution in free suspension (shown black). Classical models of spheroid growth are capable of capturing growth in free suspension to excellent accuracy, as highlighted by the least-squares fit of a Greenspan-inspired model to the free-suspension data, shown in black. Details of the fitting can be found in \cref{app: numerical details}.}
    \label{fig: helmlinger intro}
\end{figure}

\changed{Throughout our exploration of tumour growth models}, we will seek a model that exhibits a number of properties, each of which will focus on
robustly reproducing features of experimentally observed tumour dynamics and
having behaviour consistent with our understanding of tumour
growth. \changed{Before we describe these properties,} it will be helpful to
first illustrate a known impact of mechanical factors on tumour growth
dynamics. To this end, in \cref{fig: helmlinger intro} we showcase a
selection of the experimental data reported by \citet
{Helmlinger1997}, as digitised by \citet{Yan2021}; the various
growth curves correspond to differing levels of mechanical resistance
exerted on growing spheroids embedded in agarose gels of various
concentrations and, hence, stiffnesses. From these datapoints alone, it is
clear that the mechanical properties of the external medium can
significantly impact tumour growth dynamics in this system, with increasing
stiffness reducing a spheroid's capability to grow.

The simplest models of tumour growth neglect mechanical effects and assume
that growth depends only on the availability of diffusing nutrients, such as
oxygen. Such models have proven successful in reproducing the growth of
tumour spheroids in free suspension. We illustrate this by fitting a model
inspired by Greenspan's seminal model \citep{Greenspan1972} of
nutrient-limited growth to the free-suspension growth curve in \cref
{fig: helmlinger intro}; full details of the employed model are provided
later in \cref{sec: nutrient driven growth}, and the fitting process is
summarised in \cref{app: numerical details}. Whilst the mathematical model
exhibits excellent agreement with the experimental observations of \citet{Helmlinger1997} for tumour growth in free suspension, it is not
able to simultaneously fit all three datasets shown, as mechanical effects
are neglected. Motivated by this, a further aim of this \changed{tutorial} is
to formulate Greenspan's \changed{classical} approach within a
solid-mechanics framework. In particular, we will seek to capture the
phenomena reported by \citeauthor{Helmlinger1997}, with the specific goal of fitting a
mechanical model to the full range of dynamics shown in \cref{fig: helmlinger
intro}.

With this \changed{additional goal} in mind, we specify a basic requirement of
any model that we will develop: it must give rise to growth curves that are
qualitatively similar to those of \cref{fig: helmlinger intro}. More
specifically, growth curves must be monotonic and should capture a profile of
development that is approximately of exponential, linear, then saturating
character, as observed in Helmlinger et al.'s experiments and as is canonical
of tumour spheroid growth \citep{Bull2022}. Here, we define \emph
{saturating} dynamics to be those whose growth rate tends monotonically to
zero, noting that this does not guarantee that the tumour size itself is
bounded\footnote{\changed{Consider, for instance, $y = \sqrt{t}$, which has
$\mathrm{d}y/\mathrm{d}t \rightarrow 0$ but $y$ unbounded as
$t\rightarrow\infty$.}}. However, a stable non-zero steady state of tumour
size is often associated with these growth curves, the existence of which we
will \changed{also} look to guarantee analytically.

As solid mechanics will necessarily play a key role, we will also seek certain
properties that \changed{relate} to the mechanical stress within the
spheroid. A minimal such requirement is that the solid stresses in the tumour
should be bounded, so that \changed{a model will not predict that a tumour
experiences} arbitrarily large internal stresses as time increases. Whilst
this intuitive property might seem to be elementary to realise, we will show
that it does not necessarily hold \changed{in even simple cases, and
therefore} requires appropriate consideration. \changed{Finally, we will also
seek out} the ability of a model to reproduce profiles of \changed
{solid} stress that are similar to those \changed{that have been estimated}
from experimental data. In particular, we will aim to qualitatively match
profiles of \emph{residual stress}, the term given to solid stresses
accumulated in a tumour during growth that remain present when the tissue
experiences no external mechanical load. These mechanical properties, along
with the features introduced above, are summarised in \cref{tab: criteria
intro}.

\changed{In summary, in this tutorial we will seek to develop a continuum
model of avascular spheroid growth, one in which growth is regulated by
mechanical effects and nutrient availability. In attempting to realise the
noted desirable properties, we will describe and explore a number of
intermediate models, highlighting how an iterative, minimalistic approach to
model construction can provide insight into emergent behaviours and
facilitate the development of mathematical models that are simple,
interpretable, and robust. We will begin by incorporating the established
foundation of Greenspan \citep{Greenspan1972} within the framework of
morphoelasticity, striving throughout for simplicity in order to enable exploratory and analytical study.}

\begin{table}[tbhp]
    \centering
    \footnotesize
    \caption{Desirable properties of a minimal model of tumour spheroid growth. The Greenspan-inspired model used in \cref{fig: helmlinger intro}, described in \cref{sec: nutrient driven growth} and labelled `1' in this table, 
    robustly generates plausible tumour growth curves and possesses a stable steady state of tumour size, but it does not include mechanical feedback. Without further analysis, it is unclear if the model would yield bounded solid stresses or realistic residual stress profiles when cast in a solid mechanical framework.}
    \begin{tabular}{P{12mm}*{5}{|P{18mm}}}
        \tableheader\\ \hline
        \hyperref[sec: nutrient driven growth]{1} & \xmark & \checkmark & \checkmark  & -- & --
    \end{tabular}
    \label{tab: criteria intro}
\end{table}

\section{Continuum-mechanical framework}
\subsection{Geometry and set-up}
\begin{figure}[t]
    \centering
    \includegraphics[width=0.6\textwidth]{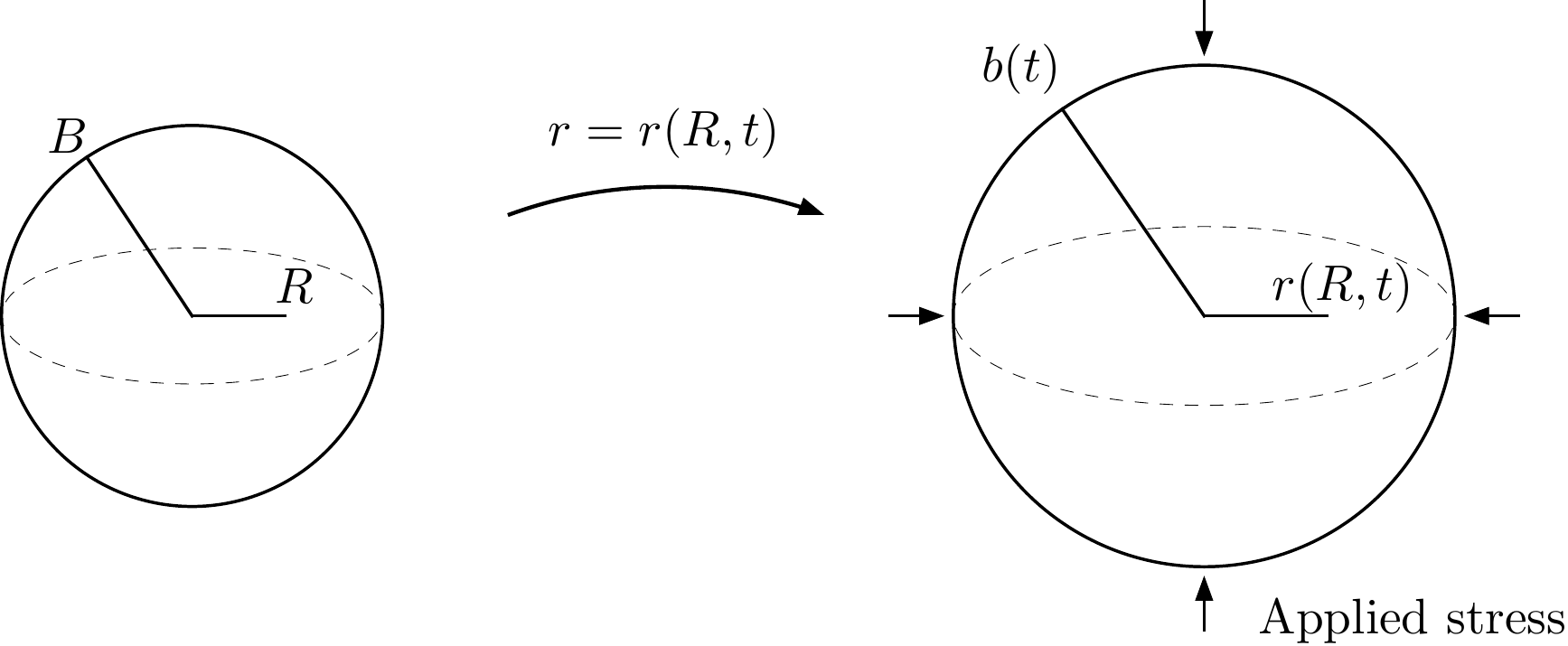}
    \caption{Geometry and set-up. The initial, stress-free configuration of the spheroid, parameterised by the Lagrangian radial coordinate $R\in[0,\initialOuterRadius]$, deforms to a loaded Eulerian configuration at time $t$ with radial coordinate $r\in[0,\outerRadius(t)]$, preserving spherical symmetry. An extracellular medium imparts a uniform compressive radial stress on the surface of the deformed spheroid.}
    \label{fig: geometry}
\end{figure}

Throughout this \changed{tutorial}, we will model a tumour spheroid as a
morphoelastic solid \citep{Goriely2017}, considering its deformation due to
the processes of growth and elastic relaxation whilst assuming strict
spherical symmetry, as illustrated in \cref{fig: geometry}. Deformations are
captured by the time-dependent gradient tensor $\tensor{F}$, which encodes
the mapping from the Lagrangian spherical coordinates $(R,\Theta,\Phi)$ of
the initial spheroid, assumed to be stress free, to the Eulerian spherical
coordinates $(r,\theta,\phi)$ that represent the deformed configuration of
the spheroid. With our assumption of spherical symmetry, the deformation
gradient, written in the orthonormal spherical bases, reads
\begin{equation}
    \tensor{F} = \pdiff{(r,\theta,\phi)}{(R,\Theta,\Phi)} = 
    \diagMat{\pdiff{r}{R}}{\frac{r}{R}}{\frac{r}{R}}\,,
\end{equation}
where $r = r(R,t)$ is a function of time $t\geq0$ and the Lagrangian radial
coordinate $R\in[0,\initialOuterRadius]$, with $r(R,0) = R$ and where
$\initialOuterRadius$ is the initial radius of the spheroid. Throughout, we
assume that the mapping between the initial and deformed configurations is
such that $\det{\tensor{F}(R,t)}>0$ for all $t\geq0$ and all $R\in[0,B]$, so
that the deformation preserves the orientation of the material and is locally
injective for all $t$. We also assume that the spheroid undergoes no
topological changes, so that $r(0,t) = 0$, and we denote the outer radius of
the deformed tumour by $\outerRadius(t)\coloneqq r(\initialOuterRadius,t)$.

\subsection{Morphoelasticity}\label{sec: morpho}
Following \citet{Goriely2017}, the central assumption of
morphoelasticity is that the tensor $\tensor{F}$ can be  multiplicatively
decomposed into two components: one that represents the growth of the
material, and another that captures the elastic response of the grown
material:
\begin{equation}
    \tensor{F} = \tensor{A}\tensor{G}\,,
\end{equation}
where $\tensor{A}$ and $\tensor{G}$ are second-order tensors that represent
the effects of elasticity and growth, respectively. Appealing to spherical
symmetry, this relation can be explicitly written as
\begin{equation}\label{eq: explicit morphoelastic decomposition}
    \diagMat{\pdiff{r}{R}}{\frac{r}{R}}{\frac{r}{R}} = \diagMat{\elasticStretch_r}{\elasticStretch_\theta}{\elasticStretch_\phi} 
    \diagMat{\growthStretch_r}{\growthStretch_\theta}{\growthStretch_\phi}\,,
\end{equation}
where $\elasticStretch_r, \elasticStretch_\theta,\elasticStretch_\phi$ and
$\growthStretch_r, \growthStretch_\theta,\growthStretch_\phi$ are
non-negative scalar elastic stretches and growth stretches, respectively. In
particular, $\growthStretch_r$ captures material growth in the radial
direction, whilst $\growthStretch_\theta\equiv\growthStretch_\phi$ encodes
circumferential growth. The equivalence between $\growthStretch_\theta$ and
$\growthStretch_\phi$ is a consequence of spherical symmetry and,
analogously, we have $\elasticStretch_\theta\equiv\elasticStretch_\phi$.
Here,  a growth stretch larger than one corresponds to the addition of
material, whilst resorption occurs when the growth stretch is less than 1
(but larger than 0).

Taking the determinant of \cref{eq: explicit morphoelastic decomposition}
yields the following quasistatic partial differential equation for the radial
coordinate
\begin{equation}\label{eq: PDE for r before incomp.}
r^2\pdiff{r}{R} =  \elasticStretch_r\elasticStretch_\theta^2\growthStretch_r\growthStretch_\theta^2R^2\,,
\end{equation}
where we have replaced $\elasticStretch_\phi$ and $\growthStretch_\phi$ by
$\elasticStretch_\theta$ and $\growthStretch_\theta$, respectively.
Henceforth, for simplicity and consistent with our goal of pursuing a minimal
model of spheroid growth, we will assume that the material is incompressible,
so that elastic deformations do not alter the volume. With this being
equivalent to imposing the condition $\det{\tensor
{A}} = \elasticStretch_r\elasticStretch_\theta^2 = 1$, \cref{eq: PDE for r
before incomp.} now takes on the simplified form
\begin{equation}
   r^2\pdiff{r}{R} = \growthStretch_r\growthStretch_\theta^2R^2\,,
\end{equation}
and elasticity no longer plays an explicit role in the instantaneous
configuration of the spheroid. In particular, given the instantaneous growth
stretches, one may readily integrate this equation, with the condition $r
(0,t)=0$, to determine $r(R,t)$, without further consideration of mechanical
effects. From this, one might be tempted to neglect mechanics altogether;
however, we will see that incorporating mechanics in an evolution law for the
growth stretches is needed in order to reproduce the properties listed
in \cref{tab: criteria intro}. 

Writing $\elasticStretch_r = \elasticStretch_\theta^{-2}$ and defining
$\elasticStretch\coloneqq\elasticStretch_\theta$, the decomposition of \cref
{eq: explicit morphoelastic decomposition} then yields
\begin{subequations}\label{eq: drdR and r over R in terms of alpha and gamma}
\begin{align}
    \pdiff{r}{R} &= \elasticStretch^{-2}\growthStretch_r\,,\label{eq: drdR in terms of alpha and gamma r}\\
    \frac{r}{R} &= \elasticStretch\growthStretch_\theta\,,\label{eq: r over R in terms of alpha and gamma theta}
\end{align}
\end{subequations}
which will later enable us to eliminate the elastic stretch $\elasticStretch$
from calculations and to transform between Lagrangian ($R$) and Eulerian
($r$) coordinates.

\subsection{Mechanics and constitutive assumptions}\label{sec: mechanics}
We will assume that the spheroid is composed of an isotropic hyperelastic
material, such that it is characterised by a strain energy function $W$. With
$W$ thereby a function of the principal stretches, which are the eigenvalues
of the right Cauchy-Green tensor $\tensor{A}\tensor{A}^T$, we may generically
write $W = W
(\elasticStretch_r, \elasticStretch_\theta, \elasticStretch_\phi)$. The
Cauchy stress tensor $\tensor{\sigma}$ is then given by
\begin{equation}
    \tensor{\sigma} = \tensor{A}\pdiff{W}{\tensor{A}} - p\eye\,,
\end{equation}
where the pressure $p$ is the Lagrange multiplier required to enforce incompressibility and the $(i,j)$\textsuperscript{th} entry of $\partial W/\partial \tensor{A}$ is defined to be the derivative of $W$ with respect to the $(j,i)$\textsuperscript{th} entry of $\tensor{A}$. In our spherically symmetric system, we write
\begin{equation}
    \tensor{\sigma} = \diagMat{\sigma_r}{\sigma_\theta}{\sigma_\theta}\,,
\end{equation}
leading to the scalar relations
\begin{subequations}
\begin{align}
    \sigma_r &= \elasticStretch_r \pdiff{W}{\elasticStretch_r} - p\,,\\
    \sigma_\theta &= \elasticStretch_\theta \pdiff{W}{\elasticStretch_\theta} - p
\end{align}
\end{subequations}
for the radial stress $\sigma_r(R,t)$ and the hoop stress $\sigma_\theta(R,t)$
. Eliminating the Lagrange multiplier and defining $\tilde{W}$ such that $W = \tilde{W}(\elasticStretch)$, we obtain the single equation
\begin{equation}\label{eq: relating sigma_theta and sigma_r}
    \sigma_\theta = \sigma_r + \frac{\elasticStretch}{2}\diff{\tilde{W}}{\elasticStretch}
\end{equation}
relating the stresses in the circumferential and radial directions. In the absence of body forces, conservation of linear momentum reads
\begin{equation}\label{eq: conservation of linear momentum}
    \div{\tensor{\sigma}} = 0\,,
\end{equation}
where the divergence operator is with respect to the Eulerian coordinate system. Using the assumed spherical symmetry and the relation of \cref{eq: relating sigma_theta and sigma_r}, \cref{eq: conservation of linear momentum} gives
\begin{equation}
    \pdiff{\sigma_r}{r} = \frac{\elasticStretch}{r}\diff{\tilde{W}}{\elasticStretch}\,.
\end{equation}
Seeking simplicity, we suppose that the tumour is a neo-Hookean material, so that
\begin{equation}
    \tilde{W} = \frac{\shearModulus}{2}\left(\elasticStretch^{-4} + 2\elasticStretch^{2} -3 \right)\,,
\end{equation}
where $\shearModulus>0$ is the material-dependent shear modulus and $\elasticStretch^{-4} + 2\elasticStretch^{2}$ is the first invariant \changed{(trace)} of the right Cauchy-Green tensor in our simplified setting. Hence, the conservation equation for the radial stress becomes
\begin{equation}\label{eq: Eulerian PDE for radial stress}
    \pdiff{\sigma_r}{r} = 2\shearModulus\frac{\elasticStretch^2 - \elasticStretch^{-4}}{r}\,.
\end{equation}
Transforming to Lagrangian coordinates using \cref{eq: drdR in terms of alpha and gamma r} and eliminating $\alpha$ via \cref{eq: r over R in terms of alpha and gamma theta}, we obtain a quasistatic Lagrangian PDE for the radial stress,
\begin{equation}\label{eq: Lagrangian PDE for radial stress}
    \pdiff{\sigma_r}{R} = 2\shearModulus\growthStretch_r\frac{r^6 - \growthStretch_\theta^6 R^6}{r^7}\,.
\end{equation}
To model resistance to expansion due to external material, we prescribe a compressive radial stress on the boundary of the spheroid that opposes growth.
Specifically, we impose
\begin{equation}\label{eq: radial stress boundary condition}
    \sigma_r(\initialOuterRadius,t) = -\springConstant \frac{\outerRadius(t) - \initialOuterRadius}{\initialOuterRadius}\,,
\end{equation}
where $\springConstant\geq0$ encodes the stiffness of the surrounding medium, though we note that generalisations of this relation are straightforward to accommodate. With this boundary condition, which reduces to growth in free suspension when $\springConstant=0$, we can integrate \cref{eq: Lagrangian PDE for radial stress} inwards from the boundary to yield $\sigma_r$, with integration being performed numerically in practice. We can then construct $\sigma_{\theta}$ via
\begin{equation}
    \sigma_{\theta} = \sigma_r + \shearModulus\left(\elasticStretch^2 - \elasticStretch^{-4}\right)\,,
\end{equation}
or, equivalently,
\begin{equation}\label{eq: hoop stress in terms of radial stress and r}
    \sigma_{\theta} = \sigma_r +  \frac{r}{2}\pdiff{\sigma_r}{r}\,.
\end{equation}

\subsection{Growth dynamics}\label{sec: growth dynamics}
It remains to specify how the growth stretches evolve with time from their common initial value of unity. Even in the case of isotropic growth, which we will assume hereafter, it is unclear how the growth of tissue should depend on the state of the tumour. Indeed, it is this dependence that we will explore and vary throughout this \changed{tutorial}, seeking a phenomenological growth law that gives rise to the canonical growth dynamics defined earlier (exponential, linear, saturating) whilst being free from unphysical behaviours. Each growth law will take the same basic form, which we state generically in terms of $\growthStretch\coloneqq\growthStretch_r=\growthStretch_{\theta}$ as
\begin{equation}\label{eq: generic growth law}
    \frac{1}{\growthStretch}\pdiff{\growthStretch}{t} = \growthRateConst f(\tensor{\sigma},c)
\end{equation}
for a fixed rate constant $\growthRateConst$ and initial condition of unity, where the function $f$ encodes the dependence of growth on the stress tensor $\tensor{\sigma}(r,t)$ and a generic diffusible nutrient with concentration $c(r,t)$. We interpret this nutrient as oxygen, as in \citet{Greenspan1972}.

Seeking to model the growth of an avascular tumour, so that the only source of nutrients is via diffusion from the outer boundary of the spheroid, we suppose that the non-negative nutrient concentration $c(r,t)$ is governed by the PDE
\begin{equation}\label{eq: nutrient diffusion equation general}
    \pdiff{c}{t} = \diffusionCoeff \lap{c} - \consumptionRate\,, \quad r \in [0,\outerRadius(t)]
\end{equation}
wherever $c$ is nonzero, where $\diffusionCoeff$ is the diffusion coefficient of the nutrient in the tumour medium and $\consumptionRate$ is the constant consumption rate of nutrient by the tissue. We impose the simple boundary condition $c(\outerRadius(t),t) = \boundaryNutrient{}$ at the surface of the tumour, which allows us to define the diffusive lengthscale $\diffusiveLengthscale\coloneqq \sqrt{\diffusionCoeff\boundaryNutrient/\consumptionRate}$ and characteristic timescale $T=1/\growthRateConst\boundaryNutrient$ of the spheroid problem. Seeking a quasistatic solution of \cref{eq: nutrient diffusion equation general} that is a function purely of the Eulerian coordinate $r$, noting that the timescales of diffusion are typically much shorter than those of biological growth, \cref{eq: nutrient diffusion equation general} reduces to 
\begin{equation}\label{eq: nutrient pde}
    \frac{1}{r^2}\pdiff{}{r}\left(r^2\pdiff{c}{r}\right) = \frac{\consumptionRate}{\diffusionCoeff}\,.
\end{equation}
At the centre of the tumour, symmetry considerations imply a no-flux condition $\partial c / \partial r (0,t) = 0$, which leads to the solution
\begin{equation}\label{eq: cI}
    c(r,t) = \frac{\lambda}{6D}\left[r^2-\outerRadius(t)^2\right] + \boundaryNutrient\,.
\end{equation}
\changed{However, as the nutrient concentration must be non-negative, this
solution is not valid if $\outerRadius(t)$ is sufficiently large, with $c$
predicted to be negative at the core of the spheroid. There are at least two
resolutions to this problem. One route simply modifies the consumption term
in the differential equation to include a dependence on the nutrient
concentration itself, with the simplest specifying that the consumption is
proportional to the concentration. This gives rise to a non-negative
solution that can be written in terms of hyperbolic functions.
Alternatively, if we suppose that \cref{eq: nutrient pde} holds whenever
$c>0$, one can obtain an appropriate piecewise solution that only differs
from \cref{eq: cI} whenever $\outerRadius
(t) > \outerRadiusThreshold \coloneqq \sqrt
{6\diffusionCoeff\boundaryNutrient/\consumptionRate} = \sqrt
{6}\diffusiveLengthscale$, which we explore in more detail in \cref{app: partially perfused dynamics}. However, both of these options give rise to tumour
dynamics that are essentially indistinct from those of \cref{eq: cI}. Hence,
we will assume that \cref{eq: cI} applies without further consideration,
seeking simplicity in the analysis that follows and implicitly considering
spheroids that are sufficiently small so as to justify this assumption. It
is straightforward, but notationally cumbersome, to pursue our analysis with
either of the alternative nutrient profiles and relax this assumption of
smallness.}

\subsection{Governing equations}
For completeness, we now state the full system of equations governing the evolution of the solid tumour, incorporating each of the assumptions detailed above:
\begin{subequations}\label{eq: governing equations}
\begin{align}
    r^2\pdiff{r}{R} &= \growthStretch^3 R^2\,,\label{eq: PDE for r}\\
    \pdiff{\sigma_r}{R} &= 2\shearModulus\gamma \frac{r^6 - \gamma^6R^6}{r^7}\,,\label{eq: PDE for sigma}\\
    \frac{1}{\growthStretch}\pdiff{\growthStretch}{t} &= \growthRateConst f(\tensor{\sigma},c)\,,\\
    c(r,t) &= \frac{\lambda}{6D}(r^2-\outerRadius(t)^2) + \boundaryNutrient\,, \label{eq: nutrient solution}
\end{align}
\end{subequations}
along with the boundary and initial conditions
\begin{equation}
    \sigma_r(\initialOuterRadius,t) = -\springConstant\frac{\outerRadius(t) - \initialOuterRadius}{\initialOuterRadius}\,,  \quad r(0,t) = 0\,, \quad r(R,0) = R\,, \quad \growthStretch(R,0) = 1\,.
\end{equation}
Integrating \cref{eq: PDE for r} in space, taking a Lagrangian time derivative, and changing integration variable yields the spatial ordinary differential equation
\begin{equation}\label{eq: general ODE for r}
    r^2\pdiff{r}{t} = 3\int\limits_0^R \growthStretch^2 \pdiff{\growthStretch}{t} \dummy{R}^2\intd{\dummy{R}} = 3\int\limits_0^r \frac{1}{\growthStretch}\pdiff{\growthStretch}{t} \dummy{r}^2\intd{\dummy{r}} = 3\growthRateConst\int\limits_0^r f(\tensor{\sigma},c) \dummy{r}^2\intd{\dummy{r}}\,.
\end{equation}

\section{In search of a realistic minimal growth law}\label{sec: exploring models}

\subsection{A minimal nutrient-limited growth model}\label{sec: nutrient driven growth}

Our first model for tumour growth draws inspiration from the classical model of \citeauthor{Greenspan1972} for nutrient-limited growth. In Greenspan's model, the rate of growth is determined by the local nutrient concentration, and thresholds of nutrient concentration determine whether a tissue is classified as proliferating, quiescent, or necrotic. Adopting this principle leads to the minimal growth law
\begin{equation}\label{eq: nutrient-driven growth law}
    \frac{1}{\growthStretch}\pdiff{\growthStretch}{t} = \growthRateConst (c - \necrosisThreshold)\,,
\end{equation}
where $\necrosisThreshold\in(0,\boundaryNutrient)$ is a fixed nutrient threshold, below which tissues reduce in size due to lack of nutrient availability. This simple form captures the notion that, given greater nutrient availability, growth will be accelerated, whilst a lack of nutrient results in cell death and decay
. In particular, we will define necrotic tissue via the nutrient threshold condition $c<\necrosisThreshold$, whilst tissues with $c\geq\necrosisThreshold$ will be referred to as proliferating or growing. Of note, we have simplified Greenspan's original model by omitting a threshold for quiescence, instead distinguishing only growing and necrotic regimes
.

\subsubsection{Steady states of growth}
\begin{figure}[t]
    \centering
    \begin{overpic}[width=0.9\textwidth]{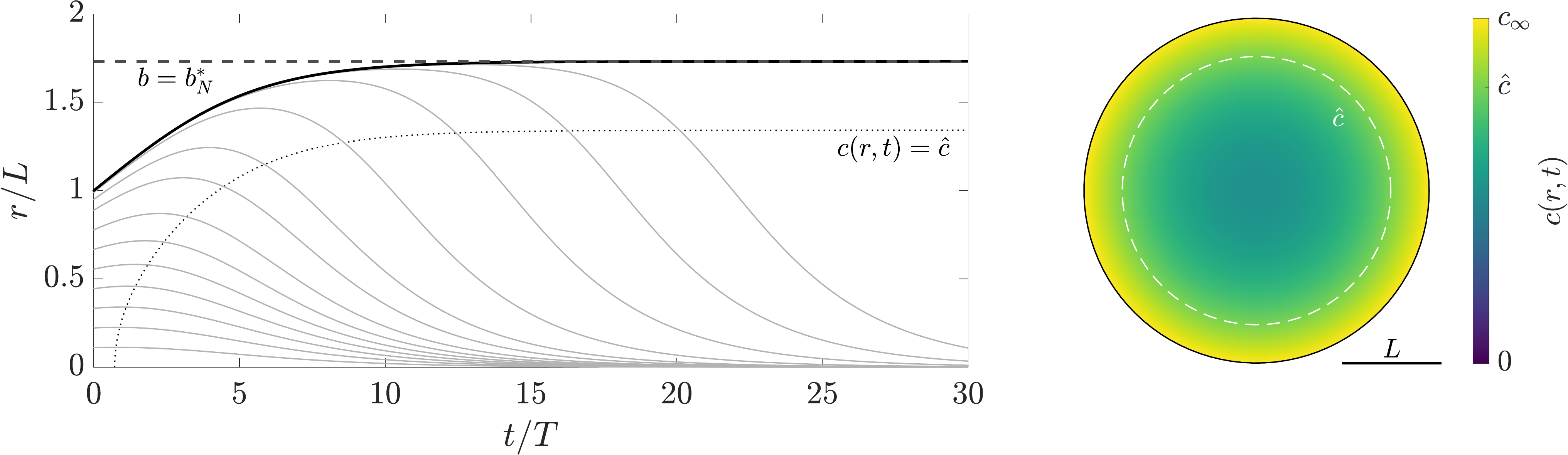}
    \put(0,31){(a)}
    \put(66,31){(b)}
    \end{overpic}
    \caption{Nutrient-driven growth of a spheroid. (a) The evolution of the tumour radius $\outerRadius(t)$ to steady state (black, solid), alongside the paths of internal material points (grey, solid).
    Material points move away from the steady outer edge of the spheroid and towards the necrotic core after the initial growth. The analytically predicted steady state for the outer edge is shown as a dotted black line and the radius at which $c(r,t)=\necrosisThreshold$ is shown as a thin dotted curve. (b) A slice through the centre of the spheroid at steady state, shaded by $c$. The threshold for necrosis, $\necrosisThreshold$, is shown as a dashed white curve. Here, $\necrosisThreshold/\boundaryNutrient = 4/5$ and $\consumptionRate\initialOuterRadius^2/\diffusionCoeff\boundaryNutrient = 1$, giving $\initialOuterRadius = \diffusiveLengthscale$ and $\ststNutrient = \sqrt{3}\diffusiveLengthscale$.}
    \label{fig: nutrient-driven growth}
\end{figure}

Inserting this growth law into our governing equations modifies \cref{eq: general ODE for r} to the simple relation
\begin{equation}\label{eq: nutrient-driven ODE for r with integral}
    r^2\pdiff{r}{t} = 3k\int\limits_0^r (c - \necrosisThreshold)\dummy{r}^2\intd{\dummy{r}}\,.
\end{equation}
With $c(r,t)$ given by \cref{eq: nutrient solution}, this integral can be evaluated explicitly, yielding the temporal evolution equation
\begin{equation}\label{eq: nutrient driven ODE for r}
    \pdiff{r}{t} = \growthRateConst r\left[ \frac{\consumptionRate}{30\diffusionCoeff}(3r^2 - 5\outerRadius^2) + \boundaryNutrient - \necrosisThreshold\right]
\end{equation}
for material points. In particular, taking $R=\initialOuterRadius$ gives an explicit ODE for the outer radius $\outerRadius(t)$ of the spheroid,
\begin{equation}\label{eq: nutrient driven outer radius ODE}
    \diff{b}{t} = \growthRateConst \outerRadius \left[ -\frac{\consumptionRate}{15\diffusionCoeff}\outerRadius^2 + \boundaryNutrient - \necrosisThreshold \right]\,.
\end{equation}
\changed{We may solve this equation to give the cumbersome but elementary explicit form
\begin{equation}\label{eq: explicit solution for outer boundary}
    \outerRadius(t) =
    \frac{\sqrt{\frac{15\diffusionCoeff(\boundaryNutrient - \necrosisThreshold)}{\consumptionRate}}}{\sqrt{1 +  \left(\frac{15\diffusionCoeff(\boundaryNutrient - \necrosisThreshold)}{\consumptionRate \initialOuterRadius^2} - 1\right)e^{-2(\boundaryNutrient - \necrosisThreshold)\growthRateConst t}}}\,,
\end{equation}
valid for a positive initial radius $\initialOuterRadius$ and $\necrosisThreshold < \boundaryNutrient$, illustrated in \cref{fig: nutrient-driven growth}a. Alternatively, a direct analysis of \cref{eq: nutrient driven outer radius ODE} yields the steady states of the dynamics, at which $\mathrm{d}b/\mathrm{d}t=0$. These steady solutions are readily seen to be $b = 0$ and $b = \ststNutrient$, where $\ststNutrient$ is defined by
\begin{equation}
  \ststNutrient \coloneqq \sqrt{\frac{15\diffusionCoeff(\boundaryNutrient - \necrosisThreshold)}{\consumptionRate}} = \sqrt{\frac{15(\boundaryNutrient - \necrosisThreshold)}{\boundaryNutrient}}\,L\,,
\end{equation} and it is straightforward to show that these states are linearly unstable and stable, respectively. Hence, so long as $\initialOuterRadius=\outerRadius(0)>0$, the tumour evolves to the state $b = \ststNutrient$, with growth limited by nutrient availability. The non-zero steady state features a necrotic core at the centre of the tumour, surrounded by a proliferative rim of tissue. The steady radius of the necrotic core, denoted by $r_{\necrosisThreshold}$, can be computed as
\begin{equation}
    r_{\necrosisThreshold} = 3\sqrt{\frac{\boundaryNutrient - \necrosisThreshold}{\boundaryNutrient}} L\,,
\end{equation}
so that $r_{\necrosisThreshold} / \ststNutrient = 3/\sqrt{15}$ and the necrotic region occupies approximately 46\% of the tumour volume, independent of the model parameters. In line with this analysis, the distribution of nutrient in the tumour at steady state is shown in \cref{fig: nutrient-driven growth}b.}

\subsubsection{Material turnover}\label{sec: nutrient-rich material turnover}
Though we have described the spheroid as being at steady state when $\outerRadius = \ststNutrient$, the tissue inside the tumour is far from idle. In particular, assuming that $\outerRadius(t)=\ststNutrient$, the spatial ODE of \cref{eq: nutrient driven ODE for r} is trivially modified to
\begin{equation}
    \pdiff{r}{t} = \growthRateConst r\left[ \frac{\consumptionRate}{30\diffusionCoeff}(3r^2 - 5(\ststNutrient)^2) + \boundaryNutrient - \necrosisThreshold\right]\,,
\end{equation}
which captures the non-steady dynamics of material within the spheroid. When viewed as an ordinary differential equation in time for $r(R,t)$ for fixed $R$, this equation admits the same steady states as that for $\outerRadius(t)$, so that the steady states are simply $r=0$ and $r = \ststNutrient$. However, for $R\in[0,\initialOuterRadius)$, the linear stability of these stationary solutions is reversed, with $r=\ststNutrient$ being unstable whilst $r=0$ is stable. Hence, material points in the interior of the spheroid move away from the outer proliferating rim and towards the central necrotic region. This behaviour, which might be expected of nutrient-limited growth, is illustrated in \cref{fig: nutrient-driven growth}a and motivates a feature of the numerical implementation later used to simulate the governing equations, as described in \cref{app: numerical details}.

\subsubsection{The growth of solid stress}\label{sec: nutrient-rich stress growth}
\begin{figure}[t]
    \centering
    \includegraphics[width=0.4\textwidth]{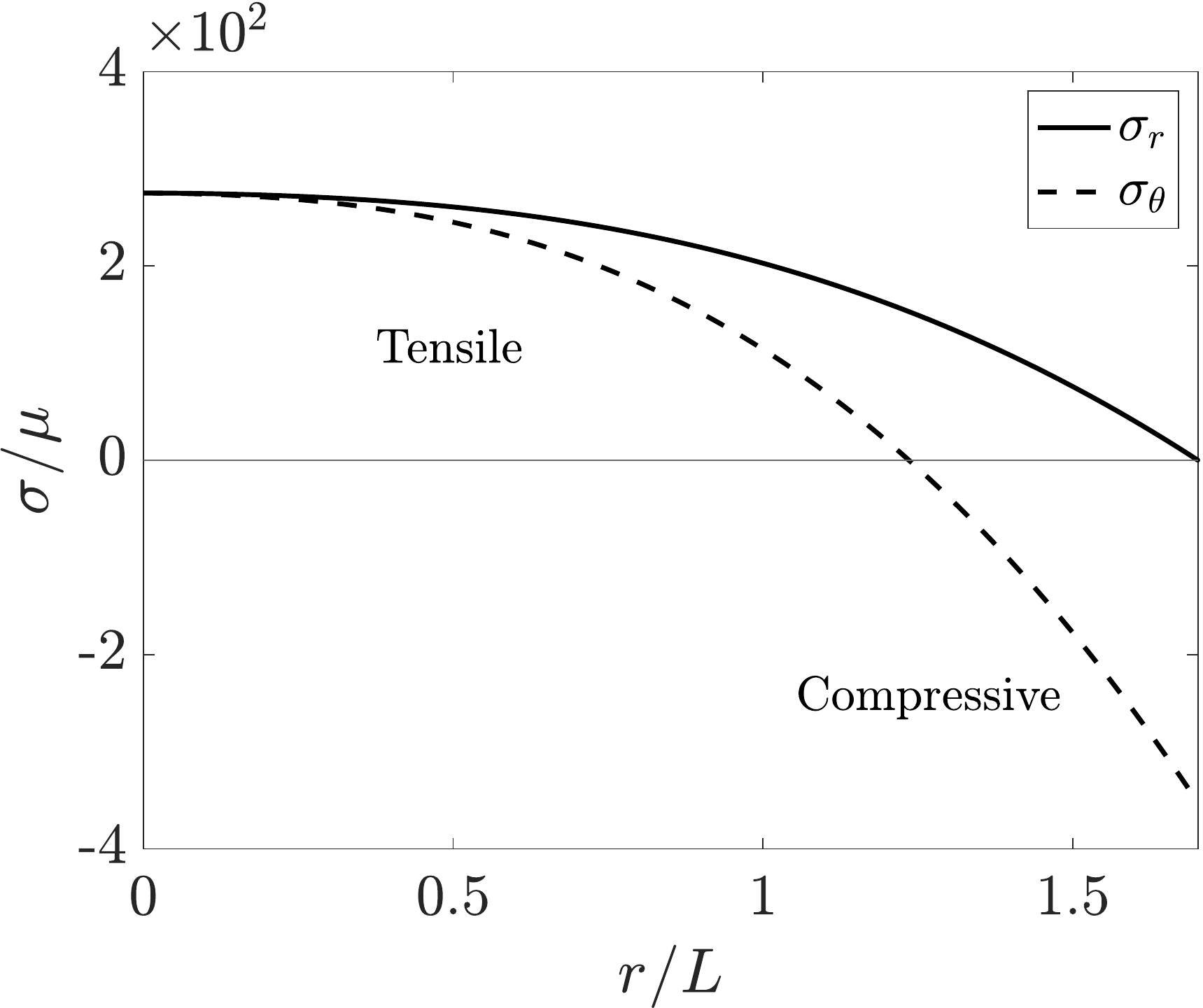}
    \caption{Accumulated stresses in a nutrient-driven spheroid. The radial and hoop stresses $\sigma_r$ and $\sigma_{\theta}$ at an instant during growth are plotted as a function of $r$, shown as solid and dashed curves, respectively. Tensile radial stress increases in magnitude towards the centre of the spheroid, whilst the hoop stress is compressive at the outer boundary of the tumour and becomes tensile at the core. Here, we have adopted the parameters of \cref{fig: nutrient-driven growth}, taken $\springConstant=0$, and sampled at time $t/T = 10$.}
    \label{fig: nutrient-driven growth: stress distribution}
\end{figure}

Whilst the simple growth law of \cref{eq: nutrient-driven growth law}, which does not incorporate mechanics, produces a plausible growth curve, we now consider whether it predicts realistic mechanical stress. Numerical solution of the governing equations of solid stress is straightforward, and the details of our implementation are discussed in \cref{app: numerical details}. In \cref{fig: nutrient-driven growth: stress distribution}, we show an illustrative stress profile at an instant during growth. Hoop stresses are compressive at the proliferating boundary and become tensile towards the centre of the tumour, whilst the radial stress is tensile throughout, with 
$\springConstant=0$ in this figure.

Turning our attention to the evolution of the solid stress, we recall that, even at a steady state of $\outerRadius(t)$, there is a constant turnover of material within the tumour. In particular, the tissue at the outer boundary of the spheroid, which remains there throughout the dynamics, grows continuously, with $c = \boundaryNutrient$ on $r=\outerRadius(t)$. Noting the simplicity that this boundary condition affords, we can explicitly write the evolution of the growth stretch at the boundary as
\begin{equation}
\at{\growthStretch}{R=\initialOuterRadius} = e^{\growthRateConst(\boundaryNutrient - \necrosisThreshold)t}\,.
\end{equation}
This then yields the elastic stretch
\begin{equation}
    \at{\elasticStretch}{R=\initialOuterRadius} = \at{\frac{r}{\growthStretch R}}{R=\initialOuterRadius} = \frac{b}{\initialOuterRadius} e^{-\growthRateConst(\boundaryNutrient - \necrosisThreshold)t}\,,
\end{equation}
which in turn allows us to evaluate the derivative of $\sigma_r$ via \cref{eq: Eulerian PDE for radial stress} as
\begin{equation}\label{eq: growing derivative of sigma_r}
    \at{\pdiff{\sigma_r}{r}}{R=\initialOuterRadius} = C_1 e^{-2\growthRateConst(\boundaryNutrient-\necrosisThreshold)t} - C_2 e^{4\growthRateConst(\boundaryNutrient-\necrosisThreshold)t}
\end{equation}
for positive constants $C_1$ and $C_2$. Perhaps surprisingly, the derivative of radial stress eventually increases in magnitude exponentially with time, driven by the exponential growth of the spheroid at the boundary, even at a steady state of the spheroid radius. Further, noting from \cref{eq: radial stress boundary condition} that $\sigma_r$ is constant on the boundary at a steady state of $\outerRadius(t)$ and 
that $\sigma_{\theta}$ is a linear combination of $\sigma_r$ and $\partial\sigma_r/\partial r$, we can also conclude that $\sigma_{\theta}$ is eventually compressive and grows exponentially in time
. Specifically, 
\begin{equation}
    \at{\sigma_{\theta}}{R=B} = -\springConstant\frac{\ststNutrient - \initialOuterRadius}{\initialOuterRadius} + \frac{\ststNutrient}{2}\left[C_1 e^{-2\growthRateConst(\boundaryNutrient-\necrosisThreshold)t} - C_2 e^{4\growthRateConst(\boundaryNutrient-\necrosisThreshold)t}\right] \sim -\frac{C_2\ststNutrient}{2}e^{4\growthRateConst(\boundaryNutrient-\necrosisThreshold)t}
\end{equation}
as $t\rightarrow\infty$. Hence, this model predicts that the solid stress in the tumour is unbounded, growing exponentially. Numerically, we observe similar behaviour in the radial stress.

The existence of this behaviour, which is confirmed to be present across parameter regimes via numerical simulations, lends itself to two distinct interpretations. One sees this prediction of ever-accumulating stress as capturing the phenomenon of spheroid shedding, whereby material is seen to break off from a tumour accompanying the destabilisation of the spherical structure \citep{Giverso2016}. In this context, one might interpret the unlimited stresses as indicative of a symmetry-breaking or topology-changing instability, though such events are beyond the reach of our framework.

Alternatively, the growing stresses might be seen instead as an
unphysical consequence of our modelling assumptions. We adopt such a viewpoint, seeking not to overreach in the interpretation of this model and its emergent dynamics. Hence, we will explore alternative growth laws that replace the minimal form of \cref{eq: nutrient-driven growth law}, with our goal being to preclude the generation of unbounded stress.

\subsection{Coupling growth to stress}\label{sec: stress-limited growth}
\subsubsection{A modified growth law}\label{sec: growth rate - stress limited}
Motivated by experimental evidence of stress-mediated regulation of cell proliferation \citep{Delarue2014,Helmlinger1997}, and building on previous works in which stress has been incorporated into the regulation of spheroid growth \citep{Ambrosi2012a,Ambrosi2017,Ciarletta2013a,Ambrosi2004}, we now couple the growth dynamics to stress. Explicitly, we pose
\begin{equation}
    \frac{1}{\growthStretch}\pdiff{\growthStretch}{t} = \growthRateConst \left\{ \begin{array}{lr}
         \stressModifier \cdot (c - \necrosisThreshold)\,, & c\geq \necrosisThreshold\,, \\
         c - \necrosisThreshold\,, &  c < \necrosisThreshold\,,
    \end{array}\right.
\end{equation}
where $\stressModifier$ is an as-yet-undefined non-negative function of $\tensor{\sigma}$ that encodes the effects of stress on growth. This form captures the intuitive principle that stress may modify the dynamics of growing tissues, 
whilst the decay of necrotic material is unaltered and consistent with Greenspan's assumptions. It is unclear how the stress modifier $\stressModifier$ should depend on the stresses experienced by the tumour: should it be a function of the radial stress, the hoop stress, or some other measure? Here, recalling that we impose a condition on the radial stress at the boundary and as a first exploration, we will suppose that $\stressModifier=\stressModifier(\sigma_r)$, so that growth is affected by the local radial stress, though we later explore an alternative.

In specifying the functional form of $\stressModifier$, we note that, unless the infimum of $n$ is zero, the stress-accumulation argument of the previous section would hold with minor modification, with this growth law therefore also leading to unbounded stresses at steady state. Seeking to avoid such an unphysical phenomenon, we  consider 
\begin{equation}\label{eq: piecewise stress modifier}
    \stressModifier(\sigma_r) = \left\{\begin{array}{lr}
        0\,, & \sigma_r\in (-\infty, \stressThreshold)\,,\\
        1 - \frac{\sigma_r}{\stressThreshold}\,, & \sigma_r \in[\stressThreshold, 0)\,,\\
        1\,, & \sigma_r\in[0,\infty)\,,
        \end{array}\right.
\end{equation}
where $\stressThreshold\leq0$ is a threshold parameter. This piecewise-linear function, which is weakly increasing in $\sigma_r$, prohibits local growth when $\sigma_r$ is sufficiently negative but has no effect when $\sigma_r$ is non-negative. This amounts to the hypothesis that compressive stresses restrict growth, with tensile stresses having no similar effect, qualitatively in line with the observations of \citet{Cheng2009,Delarue2014,Helmlinger1997}.

\subsubsection{Impact of stress-limited growth}
\begin{figure}[t]
    \centering
    \begin{overpic}[width=0.9\textwidth]{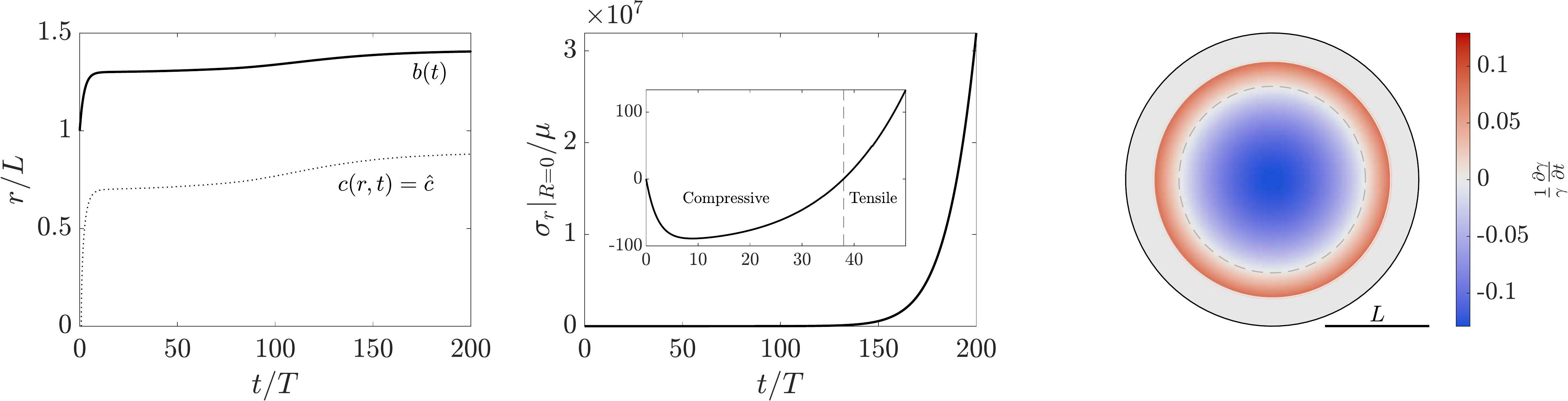}
        \put(2,26){(a)}
        \put(35,26){(b)}
        \put(69,26){(c)}
    \end{overpic}
    \caption{Development of a stress-limited spheroid. (a) The growth curve of a tumour where growth is restricted by the local radial stress, with the outer radius $\outerRadius(t)$ and the boundary of the necrotic core shown as solid and dotted curves, respectively. The tumour radius initially appears to saturate
    , though experiences an increase in growth rate around $t/T = 100$, qualitatively distinct from the nutrient-driven model of \cref{fig: nutrient-driven growth}. (b) The radial stress $\sigma_r$ at the centre of the tumour is shown as a function of time, from which an approximately exponential accumulation of stress is apparent, despite our stress-limited growth law. Inset are the dynamics at early times, with the stress initially compressive. 
    (c) The spheroid composition at $t/T=200$, shaded by growth rate, highlighting a large quiescent rim of tissue whose growth has been arrested by accumulated radial stress. Beneath this rim is a region of proliferation, shaded red, with a decaying necrotic core being present inside the grey dashed curve. Here, $\necrosisThreshold/\boundaryNutrient = 4/5$, $\springConstant/\shearModulus = 316.2$, $\stressThreshold/\shearModulus = -100$, and $\initialOuterRadius=L$.}
    \label{fig: stress-limited growth}
\end{figure}

By design, this law prohibits the growth of material that is under too much radial compression. At first glance, this might appear to solve the problem of unbounded stresses, with our argument for exponentially growing stress at the boundary no longer applying. However, with solid stress being an inherently non-local quantity, we will see that the locality of our growth law still allows for stress accumulation past any given threshold. Indeed, we exemplify such stress-limited dynamics in \cref{fig: stress-limited growth}, taking $\stressThreshold/\shearModulus=-100$ and incorporating the compressive effects of the surrounding medium by taking $\springConstant/\shearModulus=100$. The configuration of the spheroid at the final simulated time is shown in \cref{fig: stress-limited growth}c, shaded by growth rate, from which we note the presence of a quiescent outer rim of tissue whose growth has been arrested by compressive radial stress, in line with our posited growth law. However, \cref{fig: stress-limited growth}b highlights the rapid accumulation of solid radial stress in the interior of the spheroid, with the now-internal proliferative rim driving material turnover into the necrotic core. Hence, in spite of our stress-limited framework, growing stresses remain a realisable and undesirable behaviour.

This example also showcases another unrealistic consequence of this growth law. In particular, focusing on the growth curve of \cref{fig: stress-limited growth}a, we note that the dynamics near saturation are not reminiscent of typical growth profiles, with the rate of tumour growth visibly increasing, rather than decreasing, around $t/T=100$. These dynamics are due to the relaxation of accumulated stress in the proliferating rim of the tumour, the latter driven by the decay of the necrotic core and causing the corresponding increase in the stress-modulated growth rate. This leads us to question our treatment of necrosis.

\subsection{A different approach to necrosis}
\label{sec: different approach to necrosis}

\subsubsection{A modified, history-dependent growth law}\label{sec: local stress growth law with new necrosis}
A basic assumption of our growth laws thus far has been the decay of necrotic tissue, drawing inspiration from Greenspan's classical work. However, with no precise notion of material clearance present in our model, it is not clear that having such a sink of tumour mass is appropriate. The shrinkage of material also appears to be partly driving stress accumulation in the spheroid, along with the generation of atypical growth curves, as we saw in \cref{sec: stress-limited growth}. Hence, recalling our assumption of incompressibility
, we will adopt an alternative approach to necrosis, supposing instead that necrotic material simply ceases to proliferate, a condition that is permanent. This model of nutrient-starved tissues also overcomes a potential limitation of Greenspan's approach, which allows necrotic tissue to return to a proliferating state. Concretely, we pose 
\begin{equation}\label{eq: local stress growth law with new necrosis}
    \frac{1}{\growthStretch}\pdiff{\growthStretch}{t} = \growthRateConst \left\{ \begin{array}{lr}
         \stressModifier(\sigma_r) \cdot (c - \necrosisThreshold)\,, & R\geq R_N\,, \\
         0\,, &  R < R_N\,,
    \end{array}\right.
\end{equation}
where $R_N(t)$ denotes the radius of the necrotic core of the spheroid. This law prevents the growth or decay of necrotic tissue, whilst leaving the dynamics of perfused tissue unaltered from the stress-dependent law of \cref{sec: growth rate - stress limited}. Formally, the Lagrangian quantity $R_N$ is defined via the somewhat cumbersome expression
\begin{equation}
    R_N(t) = \max\left(\left\{R : c(r(R,t),t) = \necrosisThreshold\right\} \cup \left\{\sup\limits_{\tilde{t}\in[0,t)}{R_N(\tilde{t})},0\right\}\right)\,,
\end{equation}
with the intuitive interpretation that $R_N(t)$ is non-decreasing and bounded below by Greenspan's nutrient-determined necrotic radius at time $t$. This weakly increasing quantity captures the desired permanence of necrosis, whilst incorporating the principle of Greenspan's threshold-based definition. 


\begin{figure}[t]
    \centering
    \begin{overpic}[width=0.9\textwidth]{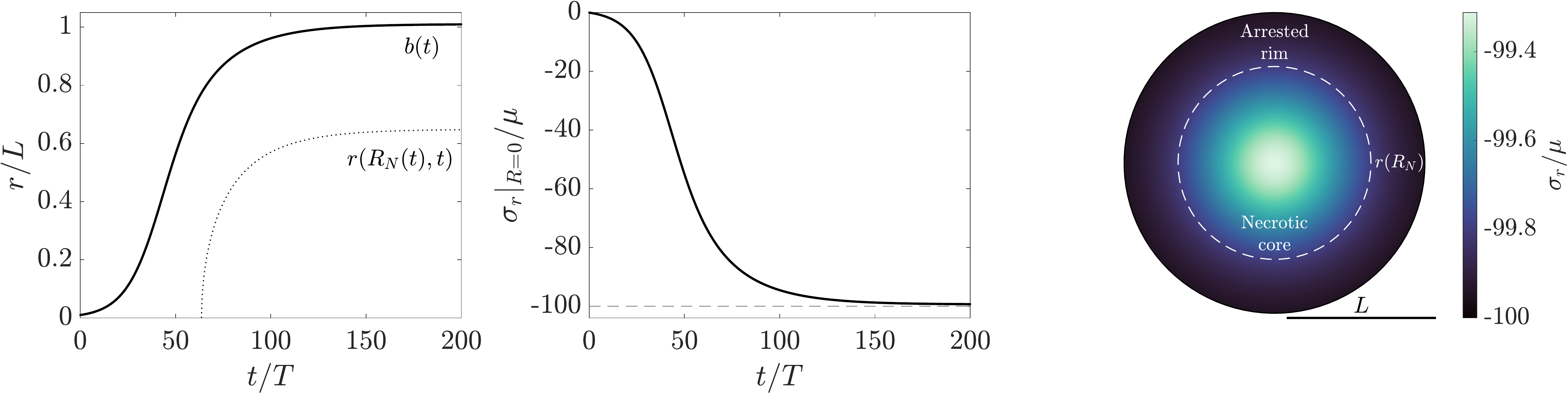}
        \put(2,26){(a)}
        \put(35,26){(b)}
        \put(69,26){(c)}
    \end{overpic}
    \caption{The growth of a stress-limited spheroid with persistent necrotic tissue. (a) The tumour growth curve and the 
    necrotic radius $r(R_N(t),t)$. 
    (b) The stress-driven saturation of radial stress at the centre of the spheroid
    . (c) The composition of the tumour and the 
    distribution of radial stress as it approaches steady state. The spheroid consists of a central necrotic region surrounded by a wide rim of stress-arrested tissue
    The dashed circle marks the current boundary of the necrotic region, with radius $r(R_N)$ at steady state. The approximately uniform radial stress distribution entails that the hoop stress is approximately equal to the radial stress, with the two quantities being indistinguishable at the resolution of these plots. Here,
    $\necrosisThreshold/\boundaryNutrient = 9/10$, $\springConstant/\shearModulus = 1$, $\stressThreshold/\shearModulus = -100$, and $\initialOuterRadius=L/100$.}
    \label{fig: growth with persistent necrotic tissue}
\end{figure}
The growth law of \cref{eq: local stress growth law with new necrosis} has an immediate consequence: the growth rate is always non-negative. Hence, the growth rate must vanish everywhere in order for the tumour to attain a steady state. This contrasts our previous explorations, which had the potential to admit a steady state of tumour radius where proliferation was balanced by the supposed decay of necrotic material. The absence of such a behaviour in our modified model results directly from the persistence of necrotic tissue, with no mechanism of material clearance being present in this simple model. However, a steady state is still attainable, resulting from the total arrest of nutrient-rich tissues by the imposed external stress. Sample growth dynamics corresponding to this model are presented in \cref{fig: growth with persistent necrotic tissue} and exemplify the desired exponential-linear-saturating growth curve. Further, the stress evolution shown in \cref{fig: growth with persistent necrotic tissue}b demonstrates the saturation of mechanical stress within the spheroid as it approaches steady state, with the entire spheroid becoming quiescent as $t\rightarrow\infty$, as illustrated in \cref{fig: growth with persistent necrotic tissue}c.

\subsubsection{The potential for unbounded evolution}
\label{sec: growth without steady state}
\begin{figure}[t]
    \centering
    \begin{overpic}[width=0.9\textwidth]{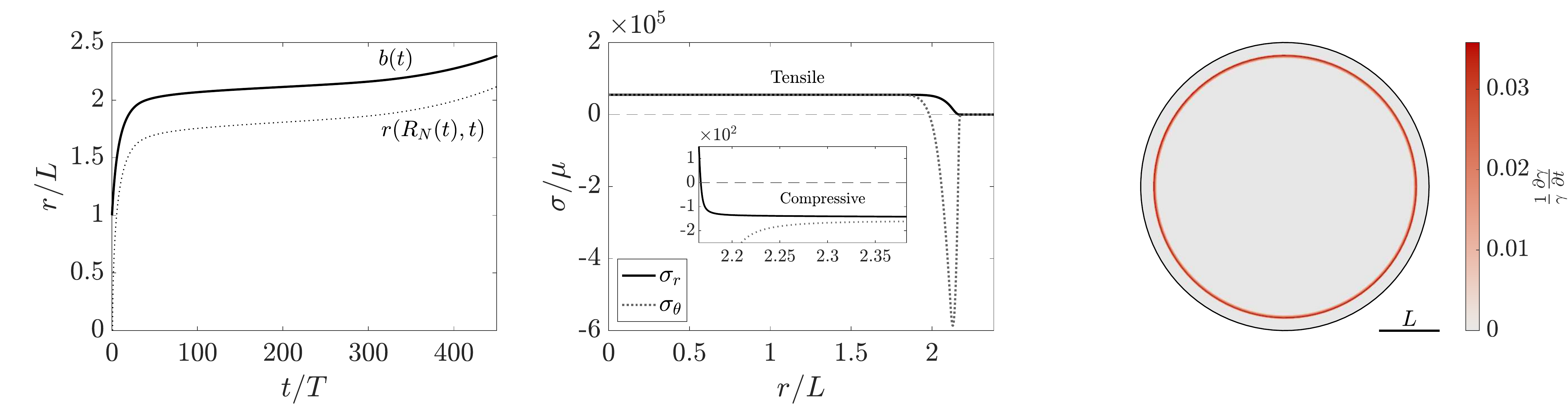}
        \put(2,26){(a)}
        \put(35,26){(b)}
        \put(69,26){(c)}
    \end{overpic}
    \caption{Stress-modulated tumour growth without a steady state
    . (a) A growth curve corresponding to a model tumour whose growth, whilst limited in theory by compressive radial stress, is accelerating after a period of apparent saturation. 
    (b) The stress profile 
    at $t/T = 450$, highlighting a region of rapid transition between high-magnitude stresses in the core and 
    compressive radial stresses at the boundary. (c) The composition of the tumour at $t/T=450$, with the narrow but persistent region of proliferation shown red, shaded by growth rate. The inner grey region corresponds to the necrotic core, whilst the outermost rim of the tumour has been arrested due to the compressive stress from the external medium. Here, $\necrosisThreshold/\boundaryNutrient = 4/5$, $\springConstant/\shearModulus = 102.4$, $\stressThreshold/\shearModulus = -100$, and $\initialOuterRadius=L$.}
    \label{fig: growth without a steady state}
\end{figure}

Though our refined model appears promising, 
the lack of a decay-driven steady state raises a new issue: it is not guaranteed that there is a non-negative steady state in all parameter regimes. Indeed, we realise such a case in \cref{fig: growth without a steady state}, where the evolution of the tumour radius deviates from the saturating profile that is typical of tumour spheroids, instead growing indefinitely. In particular, after what appears to be the onset of saturation, the tumour experiences an increase in growth rate, with a thin proliferating band of tissue growing within an outer quiescent rim. This appears to be the result of tensile radial stresses building up inside the tumour at a faster rate than can be compensated for by the linearly compressive boundary condition, yielding a narrow non-vanishing region between the necrotic core and the stress-arrested tissue where the growth rate is not identically zero. Such a persisting region of mild solid stress is visible in \cref{fig: growth without a steady state}c, with the corresponding radial and hoop stresses being illustrated in \cref{fig: growth without a steady state}b. To remove this possibility and guarantee the existence of a non-zero steady state in all parameter regimes, as per the desired properties set out in \cref{tab: criteria intro}, we further modify our growth law, focusing on the consequences of locality.

\subsection{Recovering a robust steady state}
\label{sec: recovering a robust steady state}
With the previous stress-dependent spheroid model having forgone any guarantee of a steady state in favour of stress-dependent growth and the persistence of necrotic tissue, it remains to recover the desired growth curve in all parameter regimes where external resistance is applied. In modifying our growth law appropriately, it is instructive to further consider the stress distribution in the ever-growing tumour of \cref{fig: growth without a steady state}, as illustrated in \cref{fig: growth without a steady state}b. In particular, we note that the compressive radial stress arrests growth only in the outer portion of the tumour, where $\sigma_r < \stressThreshold$, with the rest of spheroid able to grow if sufficiently perfused with nutrient. It is this \emph{locality} of the stress modulation that allows for this varied composition. Hence, 
we will modify the local nature of our stress dependence, instead modulating the growth rate by a non-local 
measure of stress. With reference to the 
biological tissues that we seek to model, such a non-local effect may be interpreted as the result of inter-cell signalling \citep{Maia2018,Aasen2016}. In particular, noting that the lengthscale of the proliferating region of tissue is dictated by the diffusion of nutrients, the diffusion of signalling molecules represents a plausible mechanism for cell-cell communication within the well-perfused regions of the spheroid.

With this interpretation in mind, we propose the following non-local growth law:
\begin{equation}\label{eq: non-local stress growth law}
    \frac{1}{\growthStretch}\pdiff{\growthStretch}{t} = \growthRateConst \left\{ \begin{array}{lr}
         \stressModifier\left(\min\limits_{\tilde{R}\in\left[0,\initialOuterRadius\right]}\{\sigma_r(\tilde{R},t),\sigma_{\theta}(\tilde{R},t)\}\right) \cdot (c - \necrosisThreshold)\,, & R\geq R_N\,, \\
         0\,, &  R < R_N\,.
    \end{array}\right.
\end{equation}
Here, in place of the local radial stress, we have adopted a measure that is both \emph{global}, in that it accounts for the stress throughout the tumour, and \emph{directionally unbiased}, with $\min_{\tilde{R}}\{\sigma_r(\tilde{R},t),\sigma_{\theta}(\tilde{R},t)\}$ being the largest magnitude compressive stress experienced by the tissue in any direction. This latter property arises as $\sigma_r$ and $\sigma_{\theta}$ are the eigenvalues of the (diagonal) stress tensor. In practice, with reference to the composition of the spheroid shown in \cref{fig: growth without a steady state}, this growth law prevents the formation of narrow bands of proliferating tissue within an arrested outer rim
.

\Cref{eq: non-local stress growth law} is identical in structure to that of \cref{sec: different approach to necrosis} and, hence, inherits the desirable properties of each of our considered models. In particular, it maintains a nutrient dependence reminiscent of Greenspan's seminal work, modified to consider a persistent core of necrotic tissue in the absence of material clearance. 
This law also enables solid stress to regulate growth within the tumour, whilst also guaranteeing the existence of a steady state.

This latter assertion warrants justification. As 
the rate of growth under our new law is non-negative, 
$r(R,t)$ is weakly increasing in time for all material points, which follows immediately from \cref{eq: general ODE for r}. In particular, the outer radius of the tumour is weakly increasing in time, with its rate of change being zero precisely at a steady state. Supposing that the tumour grows in a resistive external medium, so that $\springConstant>0$, the compressive boundary condition of \cref{eq: radial stress boundary condition} implies that $\sigma_r$ at the boundary decreases as the spheroid radius increases. Hence, assuming that the tumour does not attain a steady state by another means, the radial stress at the boundary necessarily approaches the threshold $\stressThreshold$. Thus, owing to the now-global dependence of the growth rate on the radial solid stress, tumour growth arrests throughout the entire spheroid, so that a steady state necessarily exists. This argument also places an upper bound on the tumour radius, that at which $\sigma_r(\initialOuterRadius,t) = \stressThreshold$, as specified by the compressive boundary condition of \cref{eq: radial stress boundary condition}. Of note, this reasoning applies to any form of radial stress boundary condition, subject to the assumption that it is strictly decreasing and unbounded below in the tumour radius. In the absence of a compressive external stress, i.e. $\springConstant=0$, the above argument does not apply, though an alternative argument involving the hoop stress allows us to partially recover the guarantee of a steady state in this case, discussed briefly in \cref{app: hoop stress steady state}.


\begin{figure}[t]
    \centering
    \begin{overpic}[permil,width=0.9\textwidth]{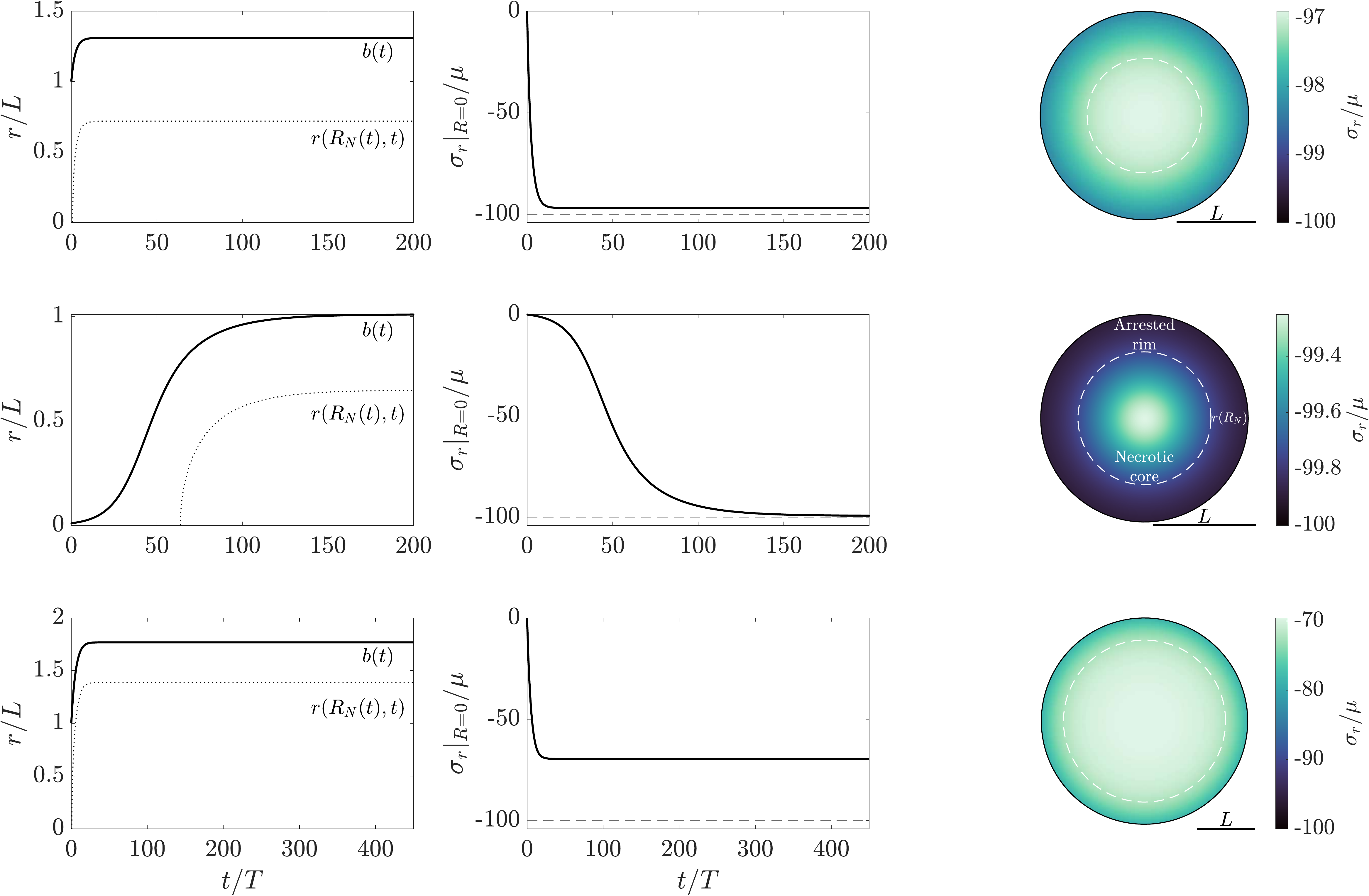}
        \put(20,668){(a)}
        \put(350,668){(b)}
        \put(690,668){(c)}
        \put(20,444){(d)}
        \put(350,444){(e)}
        \put(690,444){(f)}
        \put(20,223){(g)}
        \put(350,223){(h)}
        \put(690,223){(i)}
    \end{overpic}
    \caption{Simulated dynamics using the robust spheroid model of \cref{sec: recovering a robust steady state}. Adopting the parameters of \crefrange{fig: stress-limited growth}{fig: growth without a steady state} in turn, each row of panels reports the simulated growth curves (a,d,g), the evolution of radial stress at the centre of the tumour (b,e,h), and the distribution of radial stress at the final simulated timepoint (c,f,i). 
    In all cases we observe qualitatively plausible saturating growth curves, accompanied by saturating radial and hoop stresses (not shown)
    .}
    \label{fig: robust model examples}
\end{figure}

In order to exemplify the robustness of our proposed model and that it generates qualitatively plausible growth dynamics, we present a number of simulated spheroids in \cref{fig: robust model examples}, adopting the parameters of \crefrange{fig: stress-limited growth}{fig: growth without a steady state} in turn, two of which previously gave rise to undesirable growth curves or ceaselessly accumulating stresses. 

\subsection{Generating plausible residual stresses}
\label{sec: plausible residual stresses}
The above model satisfies all but one of the criteria set out in the Introduction, giving rise to plausible growth curves in all parameter regimes and being free from singular or inadmissible behaviours. However, 
we have not yet considered the plausibility of the generated stress profiles, except for guaranteeing that they remain bounded. In particular, we now seek to augment \cref{eq: non-local stress growth law} in order to qualitatively match experimentally determined profiles of \emph{residual stress}, i.e. stress in the absence of external loads. 
In particular, the radial component $\sigma^R_r$ of the residual Cauchy stress tensor $\tensor{\sigma}^R$ satisfies
\begin{equation}\label{eq: residual stress PDE}
     \pdiff{\sigma^R_r}{R} = 2\shearModulus\gamma \frac{r^6 - \gamma^6R^6}{r^7}\,, \quad \sigma^R_r(\initialOuterRadius,t) = 0\,,
\end{equation}
whilst the corresponding residual hoop stress $\sigma^R_{\theta}$ satisfies
\begin{equation}\label{eq: residual hoop stress}
    \sigma^R_{\theta} = \sigma^R_r +  \frac{r}{2}\pdiff{\sigma^R_r}{r}\,.
\end{equation}
In a theoretical study inspired by experimental works, it has been predicted that the residual hoop stresses in tumours may be tensile ($\sigma^R_{\theta}>0$) at the boundary of the spheroid and compressive ($\sigma^R_{\theta}<0$) further inside the tumour \citep{Stylianopoulos2012}. When such a spheroid is cut radially, the relaxation of these stresses leads to an opening of the tumour, measurements of which lead to estimates for the residual hoop stresses \citep{Stylianopoulos2012,Ambrosi2017,Guillaume2019}. As the final goal of this \changed{tutorial}, we seek to replicate key features of this qualitative profile of residual hoop stress, specifically that the residual hoop stress can be compressive within the tumour whilst also satisfying $\sigma^R_{\theta}>0$ at the surface.

To make progress, we note that the growth stretch associated with this model increases monotonically in $R$ for all times $t$, a property that they inherit from the nutrient profile
, so that $\partial\growthStretch/\partial R\geq0$ everywhere. This immediately implies $r \leq \growthStretch R$, which may be deduced via the integral formula
\begin{equation}\label{eq: r cubed}
    r(R,t)^3 = 3\int_0^R \growthStretch^3\tilde{R}^2\intd{\tilde{R}}\leq 3 \growthStretch(R,t)^3\int_0^R\tilde{R}^2\intd{\tilde{R}} = \growthStretch(R,t)^3R^3\,.
\end{equation}
The first equality in \cref{eq: r cubed} results from the integration of \cref{eq: PDE for r} in $R$, whilst the inequality stems from the monotonicity of $\growthStretch$ in $R$. Hence, \cref{eq: residual stress PDE} gives $\partial \sigma_r^R / \partial R \leq 0$, so that $\partial \sigma_r^R / \partial r \leq0$ also. Separately, noting \cref{eq: residual hoop stress} and the stress-free boundary condition on $\sigma_r^R$ at $R=\initialOuterRadius$, we can identify $\sigma^R_{\theta} = r(\partial\sigma^R_r / \partial r) / 2$ at $R=\initialOuterRadius$. Hence, the residual hoop stress $\sigma^R_{\theta}$ has the same sign as the spatial derivative of $\sigma^R_r$. Thus, we have that $\sigma^R_{\theta}\leq0$ at the boundary, so that the residual hoop stress at the surface of the spheroid is never tensile in the model of \cref{sec: recovering a robust steady state}.

From this analysis, it is clear that the monotonicity of $\growthStretch$ poses a barrier to generating a realistic profile of residual stress. However, this guarantee of monotonicity was not present in some of our earlier models, in particular those of \cref{sec: stress-limited growth} and \cref{sec: different approach to necrosis}, which each employed a coupling of the growth rate to the \emph{local} stress. Motivated by the rapidly varying profiles of hoop stress shown in \cref{fig: growth without a steady state}b, 
we now reintroduce a local stress coupling to our growth law.

In order to retain the many desirable properties of our previous model, we augment the robust growth law of \cref{sec: recovering a robust steady state} to
\begin{equation}\label{eq: combined local and nonlocal stress growth law}
    \frac{1}{\growthStretch}\pdiff{\growthStretch}{t} = \growthRateConst \left\{ \begin{array}{lr}
         \stressModifier\left(\min\limits_{\tilde{R}\in\left[0,\initialOuterRadius\right]}\{\sigma_r(\tilde{R},t),\sigma_{\theta}(\tilde{R},t)\}\right) \cdot \tilde{\stressModifier} \cdot (c - \necrosisThreshold)\,, & R\geq R_N\,, \\
         0\,, &  R < R_N\,,
    \end{array}\right.
\end{equation}
where $\tilde{\stressModifier}$ couples the growth rate to the locally experienced stress, analogous to $\stressModifier$ of \cref{sec: stress-limited growth}. This growth law thereby captures both local and non-local stress responses, in addition to enabling nutrient availability to regulate growth. Further, and by design, this growth law retains the robustness properties of the previous model whilst the local stress term allows for nuanced and, notably, non-monotonic growth stretches, as we demonstrate below. 

\begin{figure}
    \centering
    \begin{overpic}[width=0.8\textwidth]{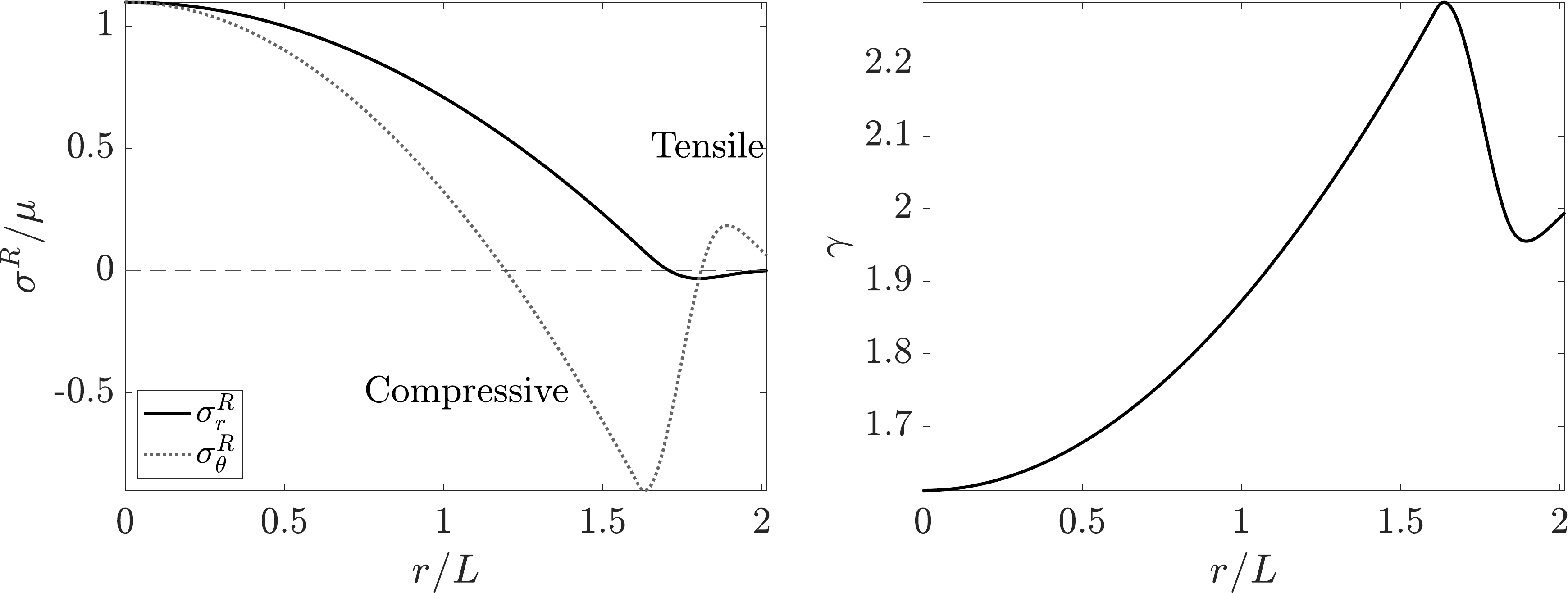}
    \put(4,40){(a)}
    \put(53,40){(b)}
    \end{overpic}
    \caption{Accumulated residual stresses and growth stretches in a tumour with both local and non-local stress dependence, following the model of \cref{sec: plausible residual stresses}. (a) The residual stress profile at steady state
    . At and near the boundary of the spheroid, the residual hoop stress is tensile, in line with experimental observations, and becomes compressive further inside the tumour. Accordingly, we see that the residual radial stress is increasing towards the boundary. (b) The growth stretches at steady state, whose non-monotonicity near the boundary is associated with the tensile residual hoop stresses shown in (a). Here, $\necrosisThreshold/\boundaryNutrient = 1/4$, $\springConstant/\shearModulus = 0.1$, $\stressThreshold/\shearModulus = -1$, $\beta=6.25$, and $\initialOuterRadius=L$.}
    \label{fig: residual stress}
\end{figure}

As with $\stressModifier$, there are countless choices of the function $\tilde{\stressModifier}$, including its argument. Here, we have taken $\tilde{\stressModifier} = \stressModifier(\beta\sigma_r)$ for a parameter $\beta>0$ that determines the relative sensitivity of the local stress response compared to the global term, so that the local response depends on the experienced radial stress, emphasising that the local and global stress responses may be wholly distinct in form. Even in this minimal case, the introduction of local stress dependence can give rise to profiles of residual stress with the desired features, though we note that such a profile is not guaranteed by this form of growth law. As an example, in \cref{fig: residual stress}a we illustrate the residual stress profiles in a simulated tumour at steady state, which displays tensile hoop stresses on the boundary and a compressive region within the tumour. Commensurate with the stress at the boundary is the non-monotonic profile of the growth stretch, as shown in \cref{fig: residual stress}b.
    
\section{Summary and evaluation}\label{sec: summary and evaluation}
Having constructed a model that \changed{satisfies the criteria set out in the Introduction}, in \cref{tab: criteria final} we summarise the iterative process that led to this final spheroid model. In this table, we identify each model by the subsection of \cref{sec: exploring models} in which it was introduced, and highlight the relevant features. 

\begin{table}[tbhp]
    \centering
    \footnotesize
    \begin{tabular}{P{12mm}*{5}{|P{18mm}}}
        \tableheader\\ \hline
        \hyperref[sec: nutrient driven growth]{1} & \xmark & \checkmark & \checkmark & \xmark & \xmark\\
        \hyperref[sec: stress-limited growth]{2} & \checkmark & \checkmark & \xmark & \xmark & \xmark\\
        \hyperref[sec: different approach to necrosis]{3} & \checkmark & \xmark & \xmark & -- & --\\
        \hyperref[sec: recovering a robust steady state]{4} & \checkmark & \checkmark$^\star$ & \checkmark & \checkmark & \xmark\\
        \hyperref[sec: plausible residual stresses]{5} & \checkmark & \checkmark$^\star$ & \checkmark & \checkmark & \checkmark$^\star$\\
        
    \end{tabular}
    \caption{Summarising model properties. Summary of the models in this study and their characteristics; a dash denotes that a property has not been explored. Models are identified by the subsection of \cref{sec: exploring models} in which they were introduced. Entries annotated with a $\star$ are conditional: models 4 and 5 have a guaranteed steady state except in a  subcase of free suspension, whilst plausible stress profiles are realisable, but not guaranteed, for model 5.}
    \label{tab: criteria final}
\end{table}

\begin{figure}[t]
    \centering
    \includegraphics[width=0.6\textwidth]{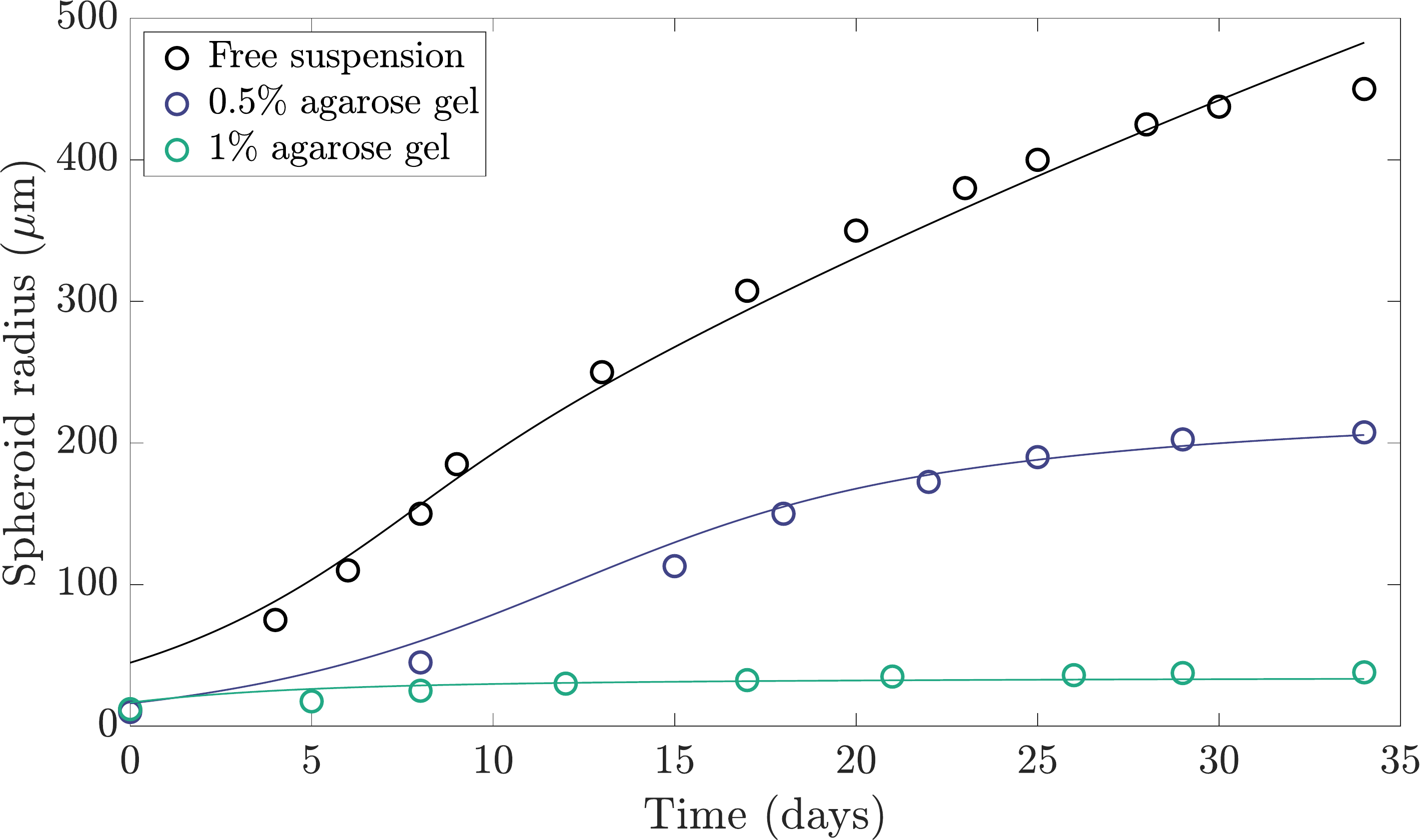}
    \caption{Fitting the refined model to the experimental data of \citet{Helmlinger1997}. The experimentally observed tumour growth dynamics described in \cref{fig: helmlinger intro} are shown in terms of the spheroid radius alongside fitted dynamics of the model of \cref{sec: recovering a robust steady state}, with the latter displayed as solid curves. \changed{The good agreement between the experimental data and the model fits highlights the ability of the constructed model to capture the phenomenon of mechanically influenced growth. In particular, we have varied only the initial conditions and external stiffness parameters between datasets during fitting the model, so that differences in dynamics can be directly attributed to these factors.}}
    \label{fig: helmlinger final}
\end{figure}

As a final evaluation of the model of \cref{sec: plausible residual stresses}, we return to the experimental data \citet{Helmlinger1997} illustrated in \cref{fig: helmlinger intro}. Recalling the inability of the minimal classical model of \citeauthor{Greenspan1972} to capture the evidenced mechanical influences on tumour growth, we now fit our final model to the three displayed growth curves simultaneously, as described in the \cref{app: numerical details}. During this fitting process, we vary only the initial condition $\initialOuterRadius$ and external stiffness parameter $\springConstant$ between the three growth curves, with all other model parameters shared. The fitted dynamics displayed in \cref{fig: helmlinger final} show good agreement between the model and experimental data. This agreement highlights the ability of our final model to capture the range of mechanically driven behaviours observed by \citet{Helmlinger1997}, with these behaviours being directly linked to the mechanical parameters of our framework. 

\section{Discussion}
In this \changed{tutorial}, we have explored a sequence of models of solid spheroid growth, ranging from simple nutrient-limited growth dynamics to more complex stress-dependent growth laws. Inspired by the classical model of \citet{Greenspan1972}, we began our exploration of tumour development by assuming a minimal threshold-based approach to determine the growth rates of tissues within the spheroid, distinguishing between proliferating and necrotic cells via the local concentration of an abstracted nutrient. The resulting growth law, when coupled to the solid framework of morphoelasticity, enabled a thorough theoretical analysis of the emerging dynamics, with simplicity affording significant tractability in this case. However, this analysis uncovered an unavoidable and unphysical behaviour in the tumour dynamics, with solid stress increasing unboundedly, even at steady states of spheroid size. Thus, we conclude that the principles of Greenspan's approach are not well-suited to the modelling of solid spheroids, at least in the context of the considered framework and without appropriate refinement. 

Following on from our analysis of a Greenspan-inspired growth law, we considered a range of refinements and modifications, introducing a dependence on both local and non-local solid stress and modifying the treatment of necrotic tissue. We demonstrated several unexpected consequences of our modifications, such as the limitless accumulation of stress in a model where stress limits and even halts tissue growth. 
This example particularly highlights how the non-locality of mechanical stress can complicate model analysis, with stress building up in non-growing regions of the spheroid due to growth elsewhere in the tumour. Nevertheless, despite the presence of this mechanical complication, we have been able to concretely reason about the behaviours of our final two models of tumour growth, concluding in both cases that the emergent dynamics necessarily reach a steady state of tumour size and solid stress, except for a subset of dynamics for growth in free suspension. This analysis highlights the benefits of employing minimal models, especially in the context of solid mechanics, with this reasoning rendered tractable by the simplicity of the morphoelastic framework in a spherical geometry. 

A further benefit of \changed{the} modelling framework \changed{explored in this tutorial} is the readiness with which we have been able to explore and experiment with different growth laws. In particular, even when analysis has not been tractable, numerical solution has been straightforward, with only the solid stresses necessitating care in order to integrate a removable singularity. In turn, this has enabled a thorough consideration of the consequences of employing solid stress as a regulator of tumour growth, from which we have seen that stress appears feasible as a factor that affects spheroid development, in support of previous experimental and theoretical works \citep{Helmlinger1997,Delarue2014,Ambrosi2004}. In particular, we have seen that a combination of local and non-local stress dependencies can give rise to robustly reasonable profiles of growth and stress, in qualitative agreement with observed tumour dynamics and estimated residual stresses. Hence, our exploration supports the broad hypothesis that tumour-environment interactions, in the particular form of mechanical stress, can be an influential and nuanced determinant of spheroid growth
. 

Complimentary to \changed{the} analysis and exploration \changed{presented here}, there \changed{are} numerous directions for future modifications to \changed{the} modelling framework and for the assessment of the role of the environment in tumour growth dynamics. One such avenue includes the relaxation of the strict assumption of spherical symmetry, which may be of pertinence to the stability of model spheroids that are highly stressed, such as those encountered in \cref{sec: nutrient driven growth}. Alternatively, there is extensive scope for the refinement of our treatment of the tumour as a single solid phase, with the potential to broaden to a poroelastic framework, as in \citet{Ambrosi2017}, or to a more general multiphase model. In particular, such an extension might include an alternative treatment of necrotic material and the inclusion of material clearance, with an explicit mechanism absent from our growth-focused exploration. Further, there is also the prospect of establishing quantitative agreement between the described models and additional experimental datasets
.

In summary, we have posed and explored a hierarchy of mechanical tumour spheroid models, exploring and iterating upon simple growth laws that draw from classical study and experimental observations. In doing so, we have seen how even simple models can give rise to unexpected and unphysical behaviours when cast in the context of solid spheroids. Seeking to preclude such eventualities, we have explored how solid stress can be used to regulate growth in phenomenological models, considering various couplings of growth to stress and evidencing the plausibility of \changed{the mechanical environment} as a driver of spheroid development. This sequence of modifications has culminated in a simple yet robust model of tumour growth that \changed{shows favourable agreement with experimental data, highlighting the potential for simple models to be more than just toy mathematical examples.}

\backmatter
\section*{Statements and Declarations}
\bmhead*{Funding}
BJW is supported by the Royal Commission for the Exhibition of 1851. GLC is supported by EPSRC and MRC Centre for Doctoral Training in Systems Approaches to Biomedical Science (grant number EP/L016044/1) and Cancer Research UK. The work of AG was supported by the Engineering and Physical Sciences Research Council grant EP/R020205/1.

\bmhead*{Code availability}
Source code and fitted parameters are freely available at \url{https://gitlab.com/bjwalker/morphoelastic-tumour}.

\bmhead*{Competing Interests}
The authors declare no competing interests.

\begin{appendices}
\crefalias{section}{appendix}

\section{Numerical implementation}\label{app: numerical details}
We numerically solve the governing equations of tumour evolution of \cref{eq: governing equations} via the method of lines in MATLAB\textsuperscript{\textregistered}, exploiting the quasistatic nature of the tumour dynamics. At each timestep, the instantaneous configuration, solid stresses, nutrient concentration, and growth rate are computed from the current growth stretches, which are then advanced in time by a forward Euler scheme. Our implementation was validated against the analytical conclusions of \cref{sec: nutrient driven growth}, with numerical convergence established for all results presented in this manuscript. Standard non-linear least-squares fitting to tumour growth curves was performed using \texttt{lsqnonlin} in MATLAB\textsuperscript{\textregistered}, which identified local minima of the appropriate residual. We also detail two notable aspects of the implementation below, with source code, fitting routines, and fitted parameters freely available at \url{https://gitlab.com/bjwalker/morphoelastic-tumour}.

\subsection{Adaptive discretisation}
The analysis of \cref{sec: nutrient driven growth} and \cref{app: partially perfused dynamics} predicts that material points will move towards the centre of nutrient-driven spheroids with a decaying necrotic core, with the exception of those precisely on the surface of the tumour. In the context of a computational implementation, this entails that any fixed discretisation of the Lagrangian domain will eventually become unsuited to capturing the entire Eulerian domain with sufficient resolution. This is illustrated in both \cref{fig: nutrient-driven growth}a and \cref{fig: partially perfused nutrient-limited growth}a, with $r(R,t)\rightarrow0$ as $t\rightarrow\infty$ for $R\in[0,\initialOuterRadius)$. To overcome this numerical issue, we rediscretise the Lagrangian domain at each timestep so that it corresponds to a uniform discretisation of the Eulerian domain, noting that $r(R,t)$ is a bijective mapping between the two domains at any time $t$.

However, this remeshing procedure results in the discrete points of the Lagrangian domain being clustered close to $R=\initialOuterRadius$, to the point that these quantities rapidly become indistinguishable when using double precision arithmetic. We mitigate this numerical limitation by casting the governing equations of \cref{eq: governing equations} in terms of a shifted Lagrangian variable, $\tilde{R}\coloneqq R-\initialOuterRadius$, with values close to zero being better distinguished in fixed-precision arithmetic than those near $\initialOuterRadius>0$.

\subsection{Computing radial stress}
\Cref{eq: PDE for sigma}, which gives the Lagrangian derivative of $\sigma_r$, has a removable singularity at $r=R=0$ that presents a significant numerical challenge. To circumvent this when integrating \cref{eq: PDE for sigma} numerically via the trapezium rule, we replace this expression with its two-term Taylor expansion about $R=0$ when $R/\initialOuterRadius<1/20$. This expansion is calculated by first computing higher-order expansions of $r(R,t)$ and $\growthStretch(R,t)$, noting once more that 
\begin{equation}
    r(R,t)^3 = 3\int\limits_0^R\growthStretch^3\tilde{R}^2\intd{\tilde{R}}
\end{equation}
from \cref{eq: PDE for r}. Writing the expansion of $\growthStretch$ with respect to $R$-derivatives of $\gamma$, after significant calculation this leads to
\begin{equation}
    2\shearModulus\growthStretch\frac{r^6 - \growthStretch^6 R^6}{r^7} \sim -3\shearModulus\frac{\growthStretch'}{\growthStretch} + \shearModulus\left[\frac{3}{40}\left(\frac{\growthStretch'}{\growthStretch}\right)^2 - \frac{12}{5}\frac{\growthStretch''}{\growthStretch}\right]R
\end{equation}
as $R\rightarrow0$. Here, a prime denotes a derivative with respect to $R$, with $\growthStretch$ and its derivatives being evaluated at $R=0$. Derivatives of $\growthStretch$ are estimated numerically with a second-order finite-difference scheme. The validity of this expansion and the resulting numerical scheme is confirmed by comparison with numerical solutions computed with high-resolution discretisations.

\section{Partially perfused nutrient-limited dynamics}\label{app: partially perfused dynamics}

\begin{figure}
    \centering
    \includegraphics[width=0.4\textwidth]{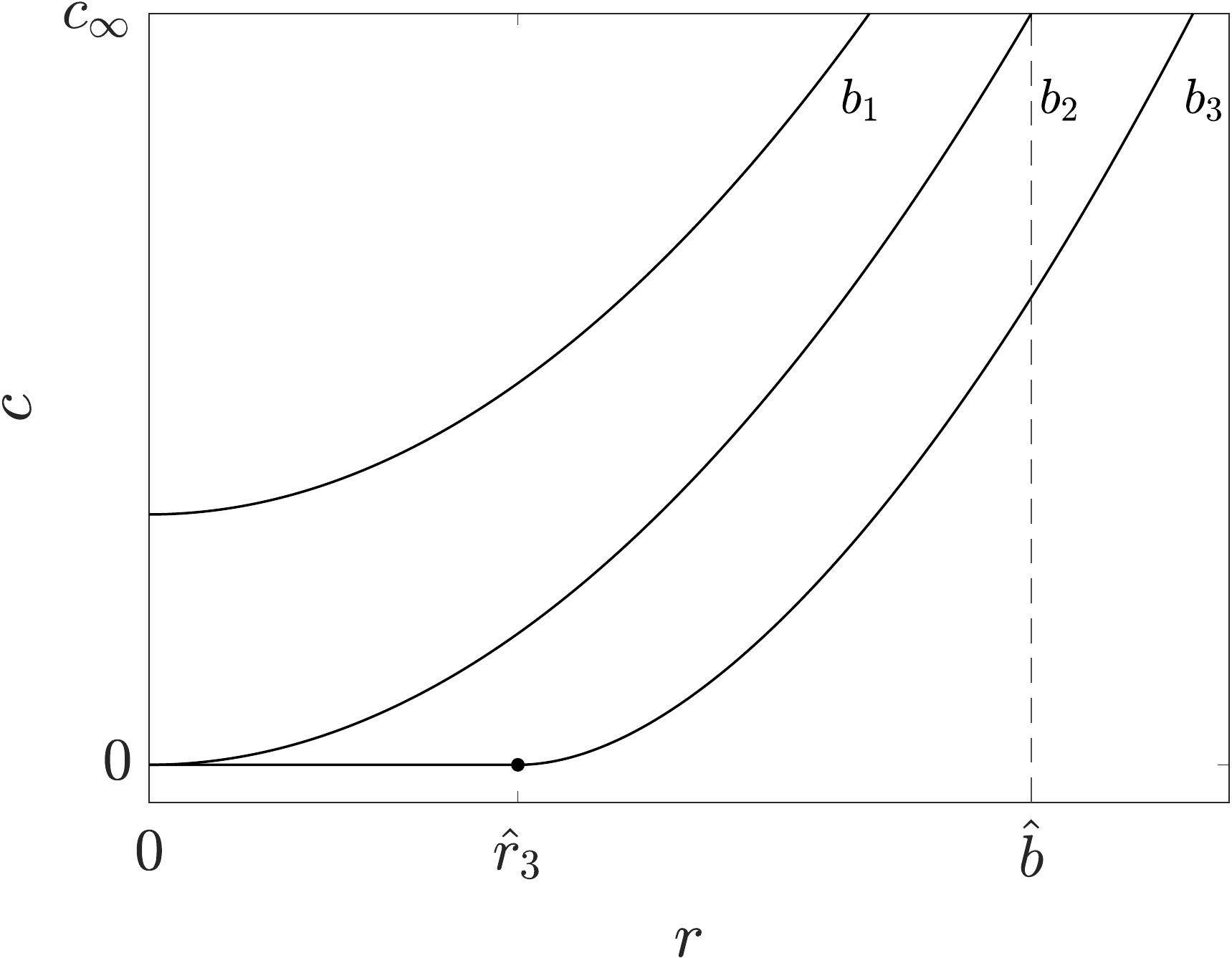}
    \caption{The nutrient concentration in a model spheroid. Sample quasistatic nutrient profiles $c$ for various $\outerRadius\in\{\outerRadius_1,\outerRadius_2,\outerRadius_3\}$ are plotted as solid black curves, which individually show the decrease in nutrient from $r=\outerRadius$ to the centre of the tumour. Here, $\outerRadius_1 < \outerRadiusThreshold$, so that $c=\cI$, $\outerRadius_2 = \outerRadiusThreshold$, where $c = \cI = \cII$, and $\outerRadius_3 > \outerRadiusThreshold$, so that $c = \cII$. For $\outerRadius = \outerRadius_3$, the threshold radius for zero nutrient concentration $\rThreshold$ is shown as a black circle and labelled $\rThreshold_3$.}
    \label{nutrient concentration}
\end{figure}

As noted in \cref{sec: growth dynamics}, larger spheroids can give rise to a solution for the nutrient concentration that differs from the expression used in the main text (see \cref{eq: cI}), with the concentration becoming zero near the tumour centre. In such a \emph{partially perfused} case, which occurs whenever $\outerRadius(t) > \outerRadiusThreshold = \sqrt{6\diffusionCoeff\boundaryNutrient/\consumptionRate} = \sqrt{6}\diffusiveLengthscale$, the solution for the nutrient concentration is instead given by
\begin{equation}\label{eq: cII}
    c(r,t) = \cII(r,t) \coloneqq \left\{\begin{array}{lr}
        0\,, & r\in[0,\rThreshold)\,, \\
        \frac{\consumptionRate}{6\diffusionCoeff}r^2 
        +
       \frac{\consumptionRate\rThreshold^3}{3D}\frac{1}{r}
        + 
        \frac{\consumptionRate(2\rThreshold^3 + \outerRadius^3)}{6\diffusionCoeff\outerRadius} + \boundaryNutrient\,, & r\in[\rThreshold,\outerRadius(t)]\,,
    \end{array}\right.
\end{equation}
where $\rThreshold$ is the smallest positive root of the polynomial
\begin{equation}\label{eq: polynomial for rThreshold}
    \lambda[2\rThreshold^3 - 3\rThreshold^2\outerRadius(t) + \outerRadius(t)^3] - 6\outerRadius(t)\diffusionCoeff\boundaryNutrient\,.
\end{equation}
This expression for the nutrient concentration satisfies both $\partial c / \partial r (\rThreshold,t) = 0$ and $c(\rThreshold,t)=0$, along with the outer boundary condition $c(\outerRadius(t),t)=\boundaryNutrient$. The overall solution for $c$ can now be succinctly written as
\begin{equation}
    c(r,t) = \left\{\begin{array}{lr}
        \cI(r,t)\,, & \outerRadius(t)\leq \outerRadiusThreshold\,,\\
        \cII(r,t)\,, & \outerRadius(t) > \outerRadiusThreshold\,,
    \end{array}\right.
\end{equation}
where $\cI$ denotes the solution of \cref{eq: cI} in the main text. This elementary but notationally cumbersome solution is illustrated in \cref{nutrient concentration}.

\subsection{Steady states of growth}
In the partially perfused regime, adopting the minimal nutrient-driven growth law of \cref{sec: nutrient driven growth}, \cref{eq: nutrient-driven ODE for r with integral} is modified to
\begin{equation}\label{eq: free suspension partially perfused ODE for r with integral}
    r^2\pdiff{r}{t} = 3k\int\limits_0^r (\cII - \necrosisThreshold)\dummy{r}^2\intd{\dummy{r}}\,.
\end{equation}
For $r<\rThreshold$, recalling the expression for $\cII$ of \cref{eq: cII}, it is clear that this spatial ODE results in movement towards the centre of the spheroid, whilst, for $r>\rThreshold$, there is the possibility of a non-zero steady state, with growth balancing out necrotic decay. For given system parameters, this steady state is straightforward to compute numerically, though presents a challenge for analytical solution due to the presence of the threshold radius $\rThreshold$, which depends on $\outerRadius(t)$. However, we can make simple progress towards a steady-state solution for $\outerRadius(t)$ under the assumption that the tumour is large, by which we mean that $\outerRadius(t)$ is larger than the other lengthscales in the problem.

To see this, we first note that $\outerRadius - \rThreshold$ represents the distance beyond which the nutrient at the boundary can no longer penetrate the tissue, a quantity that should be approximately independent of the size of the tumour when the spheroid is large. Hence, posing the asymptotic ansatz $\rThreshold \sim \outerRadius(t) - \rThreshold_0 + o(1)$ as $\outerRadius(t)\rightarrow\infty$ for some as-yet-undetermined constant $\rThreshold_0$, solving for the roots of \cref{eq: polynomial for rThreshold} gives
\begin{equation}
    (3\consumptionRate\rThreshold_0^2 - 6\diffusionCoeff\boundaryNutrient)\outerRadius(t) = 0
\end{equation}
to leading order, which simply gives $\rThreshold_0 = \sqrt{2\diffusionCoeff\boundaryNutrient/\consumptionRate} =  \sqrt{2}L$. Hence, for large spheroids, we have that $\rThreshold \sim \outerRadius(t) - \sqrt{2}L$. Inserting this approximate expression into \cref{eq: cII}, setting $r=\initialOuterRadius(t)$, and evaluating the integral of \cref{eq: free suspension partially perfused ODE for r with integral} yields
\begin{equation}
    \outerRadius^2\diff{\outerRadius}{t} = -15\growthRateConst\necrosisThreshold\outerRadius^3\left[\necrosisThreshold\outerRadius - \sqrt{2}L\boundaryNutrient  + o(1)\right]\,,
\end{equation}
which has leading-order steady states $\outerRadius = 0$ and
\begin{equation}
    \outerRadius = \ststNutrient \coloneqq \frac{\sqrt{2}L\boundaryNutrient}{\necrosisThreshold}\,,
\end{equation} redefining the nutrient-limited steady state in this context for notational convenience. It is clear that only the non-zero steady state can be consistent with the asymptotic analysis, so that the only admissible steady state scales with $\boundaryNutrient/\necrosisThreshold$, which may readily become large for sufficiently low thresholds for tissue necrosis. Further, inspecting the form of the approximate dynamics highlights that this steady state is linearly stable, consistent with the negative feedback present between tumour size and nutrient availability within the spheroid, the balance of which gives rise to this nutrient-limited steady state. 

This steady-state solution also gives the approximate steady-state value of $\rThreshold$, which we see is related to the system parameters simply as $\rThreshold \sim \sqrt{2}L(\boundaryNutrient/\necrosisThreshold - 1)$. Approximating the radius of the necrotic core by $\rThreshold$, valid in the large-$\ststNutrient$ limit, the fraction of the tumour volume occupied by the necrotic core is approximately
\begin{equation}
    \left(\frac{\rThreshold}{\ststNutrient}\right)^3 \sim \left(1 - \frac{\necrosisThreshold}{\boundaryNutrient}\right)^3\,.
\end{equation}
Perhaps counter-intuitively, a decrease in the necrosis threshold results in an increase in the proportion of necrotic tissue at steady state. However, this is easily reconciled upon noting that the tumour radius at steady state is greatly increased by the change in this parameter, whilst the perfused rim of the spheroid remains at an  approximately constant thickness.

\begin{figure}
    \centering
    \begin{overpic}[width=0.9\textwidth]{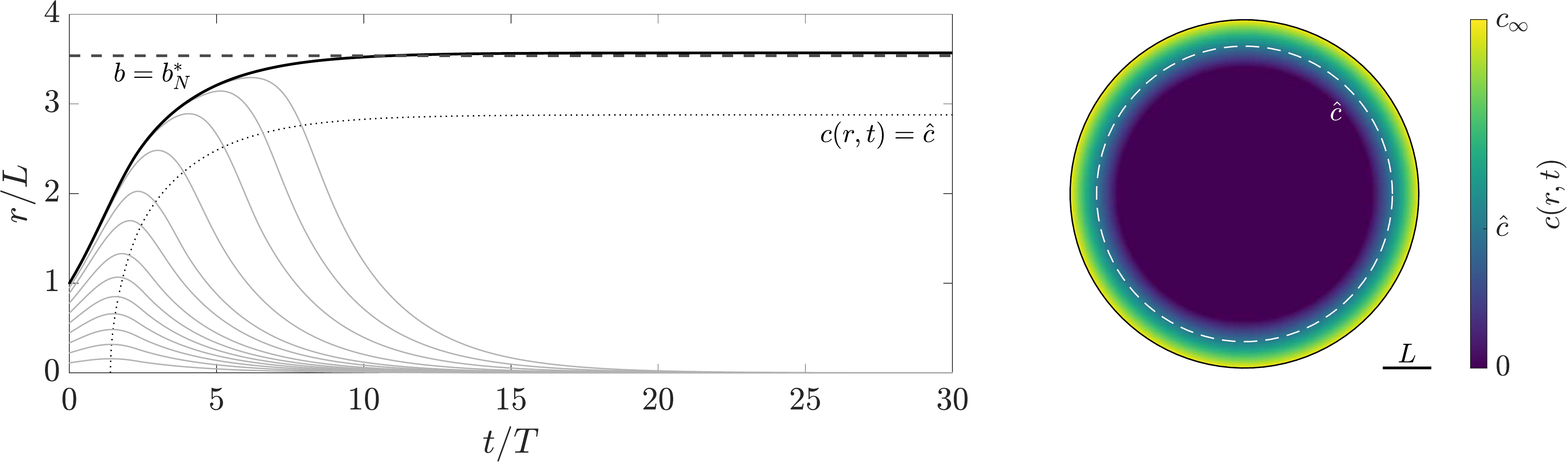}
    \put(0,31){(a)}
    \put(66,31){(b)}
    \end{overpic}
    \caption{Partially perfused nutrient-driven growth of a spheroid. (a) The evolution of the tumour radius $\outerRadius(t)$ to steady state is shown as a solid black curve, alongside the paths of material points, which are shown as solid grey curves and correspond to $R\in[0,(1-10^{-5})B]$. As in \cref{sec: nutrient driven growth}, after initial growth, material points move away from the steady outer edge of the spheroid and towards the necrotic core. The analytically predicted steady state, which here is only a leading-order approximation, is shown as a dashed black line, from which we note remarkable agreement between the numerical and asymptotic solutions even when $\outerRadius(t)$ is not large. The radius at which $c(r,t)=\necrosisThreshold$ is shown as a thin dotted curve. (b) A slice through the centre of the spheroid at steady state, shaded according to the nutrient concentration $c$, which highlights a key difference between these partially perfused dynamics and the nutrient-rich evolution of \cref{fig: nutrient-driven growth}. The threshold for necrosis, $\necrosisThreshold$, is shown as a dashed white curve, inside which the tissue shrinks in response to lack of nutrient. Here, we have taken parameters such that $\necrosisThreshold/\boundaryNutrient = 2/5$ and $\initialOuterRadius=L$. For these parameter choices, the leading-order approximation to the steady state is $\ststNutrient \sim 5\diffusiveLengthscale/\sqrt{2}$.}
    \label{fig: partially perfused nutrient-limited growth}
\end{figure}

In order to illustrate the validity of our approximate solution for the steady dynamics, particularly the tumour radius at steady state, we numerically solve the exact ODEs governing the evolution of material points in this partially perfused tumour, showcasing a sample exploration in \cref{fig: partially perfused nutrient-limited growth}. Remarkably, even for the $\outerRadius(t)< 4L$ attained in this example, we see very good agreement between the asymptotically predicted steady state $\ststNutrient$ and the numerically computed evolution, with agreement naturally improving in parameter regimes with larger steady states (i.e., those with reduced $\necrosisThreshold/\boundaryNutrient$). From \cref{fig: partially perfused nutrient-limited growth}b, we note the distinct composition of the spheroid at steady state when compared to the nutrient-rich tumour of \cref{fig: nutrient-driven growth}, with a large region devoid of nutrient present in this case and the necrotic region occupying a larger proportion of the spheroid.

\begin{figure}
    \centering
    \includegraphics[width=0.6\textwidth]{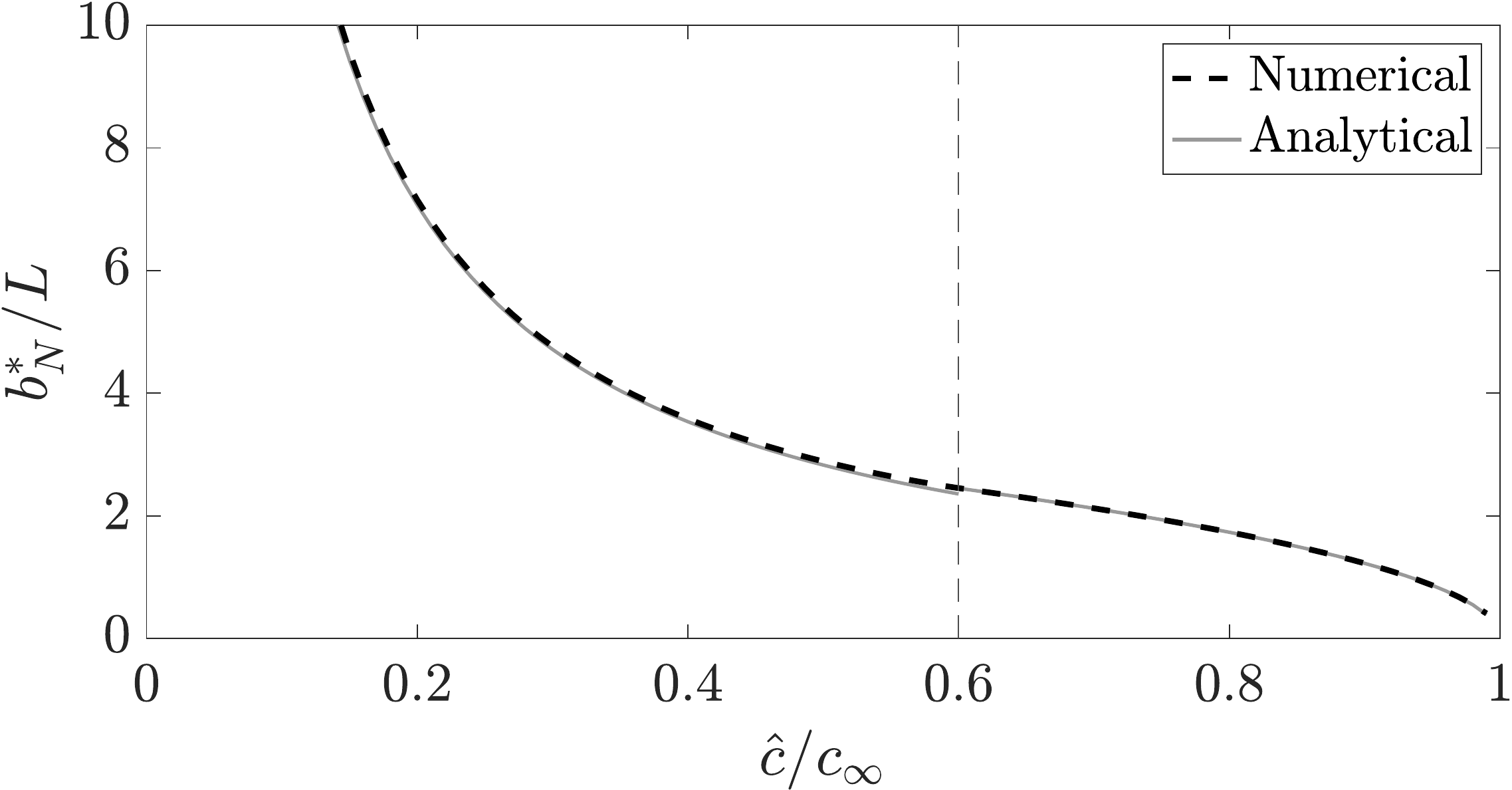}
    \caption{The non-zero steady states of nutrient-driven spheroid growth. Normalised by the diffusive lengthscale $L$, the numerically computed steady state $\ststNutrient$ of spheroid growth in the absence of mechanical effects is shown as a dashed black curve as a function of $\necrosisThreshold / \boundaryNutrient$. The analytically predicted steady states are shown as solid grey curves, with the exact result of the analysis of \cref{sec: nutrient driven growth} shown for $\necrosisThreshold / \boundaryNutrient \geq 3/5$ and the asymptotic result of \cref{app: partially perfused dynamics} shown for $\necrosisThreshold/\boundaryNutrient<3/5$. Though derived for large steady states, the asymptotic results displays a remarkable accuracy even for states of moderate magnitude. Spheroid dynamics were simulated until a time $t/T = 100$ with $\initialOuterRadius = L$, at which point the dynamics had converged to the steady state at the resolution of this figure.}
    \label{fig: nutrient-driven steady states}
\end{figure}

Additionally, with the identified  steady states having been functions of only the ratio $\necrosisThreshold/\boundaryNutrient$ and the diffusive lengthscale $L$, we illustrate the normalised non-zero steady states across the admissible range of $\necrosisThreshold/\boundaryNutrient$ in \cref{fig: nutrient-driven steady states}, including the asymptotic results of this section, the exact results of \cref{sec: nutrient driven growth}, and numerically computed steady states. From this, we observe precise agreement between the numerical results and the expression of \cref{sec: nutrient driven growth}, valid for $\outerRadius\leq\outerRadiusThreshold$, whilst the asymptotic results of this section are seen to retain substantial accuracy even for moderately sized steady states, which lie outside the regime of asymptotic validity.

\subsection{Material turnover}
\Cref{fig: partially perfused nutrient-limited growth}a also highlights the motion of material points within the spheroid, which approach the centre of the necrotic core of the tumour as the spheroid approaches steady state, as in the case of \cref{sec: nutrient driven growth}. Here, we can make simple progress in determining the evolution of much of the material in the tumour at steady state, noting that the majority of the tissue lies in the central region where $c = 0$, with $r(R,t)\leq\rThreshold$. In this region, the evolution of material points is governed by the simple spatial ordinary differential equation
\begin{equation}
    r^2\pdiff{r}{t} = -\growthRateConst\necrosisThreshold r^3\,,
\end{equation}
whose solution is exponential decay towards $r=0$ at a rate $\growthRateConst\necrosisThreshold$. For material points lying above the threshold $\rThreshold$, the analysis is once again complicated by the presence of the $\rThreshold$ term implicitly defined by \cref{eq: free suspension partially perfused ODE for r with integral}. However, for large $r$, one can once again make asymptotic progress, leading to the conclusion that $r(R,t)$ also has a steady solution $r = \ststNutrient$, though this state is linearly unstable for $R\in[0,B)$, which mirrors the analysis of the simpler, nutrient-rich case of \cref{sec: nutrient driven growth}.

\section{Recovering a steady state without external compression}\label{app: hoop stress steady state}

Consider the model of \cref{sec: recovering a robust steady state} when $\springConstant=0$, so that no external compression is applied.
Suppose that no steady state is reached and that $\stressModifier$ of \cref{eq: non-local stress growth law} attains a strictly positive minimum during the dynamics, so that $\stressModifier\geq \stressModifier_{m}>0$ for some fixed $\stressModifier_{m}$ and $\sigma_r$ and $\sigma_{\theta}$ are bounded below. Then, bounding and integrating the growth law of \cref{eq: non-local stress growth law} evaluated at $R = \initialOuterRadius$ gives the inequality
\begin{equation}
    \at{\growthStretch}{R=\initialOuterRadius} \geq e^{\growthRateConst\stressModifier_{m}^2(\boundaryNutrient - \necrosisThreshold)t}\,,
\end{equation}
so that
\begin{equation}
    \at{\elasticStretch}{R=\initialOuterRadius} \leq \frac{\outerRadius(t)}{\initialOuterRadius}e^{-\growthRateConst\stressModifier_{m}^2(\boundaryNutrient - \necrosisThreshold)t}\,,
\end{equation}
where we are mirroring the analysis of \cref{sec: nutrient-rich stress growth}. \Cref{eq: Eulerian PDE for radial stress} then allows us to bound the derivative of radial stress as
\begin{equation}
    \at{\pdiff{\sigma_r}{r}}{R=\initialOuterRadius} \leq D_1 b(t) e^{-2\growthRateConst\stressModifier_{m}^2(\boundaryNutrient - \necrosisThreshold)t} - D_2 b(t)^{-5} e^{4\growthRateConst\stressModifier_{m}^2(\boundaryNutrient - \necrosisThreshold)t}\,,
\end{equation}
where $D_1$ and $D_2$ are positive constants. Substituting this inequality into \cref{eq: hoop stress in terms of radial stress and r} and noting that $\at{\sigma_r=0}{R = \initialOuterRadius}$ when $\springConstant=0$, we have
\begin{equation}\label{eq: app: growing hoop stress}
    \at{\pdiff{\sigma_{\theta}}{r}}{R=\initialOuterRadius} \leq \frac{D_1 b(t)^2}{2} e^{-2\growthRateConst\stressModifier_{m}^2(\boundaryNutrient - \necrosisThreshold)t} - \frac{D_2 b(t)^{-4}}{2} e^{4\growthRateConst\stressModifier_{m}^2(\boundaryNutrient - \necrosisThreshold)t}\,.
\end{equation}
As the growth of large spheroids is limited by the diffusive lengthscale $L$, it is clear that $b(t)$ grows sub-exponentially in time for large spheroids. Hence, recalling that we are assuming that growth is unbounded, the hoop stress at the boundary is dominated by the growing exponential in the second term of \cref{eq: app: growing hoop stress}, so that $\at{\sigma_{\theta}}{R=\initialOuterRadius}\rightarrow -\infty$ and, consequently, $\stressModifier$ becomes zero at finite time. This contradicts our supposition, so that the tumour must evolve to a steady state, subject to the caveat that our argument does not apply to dynamics where $\stressModifier\rightarrow0$ but $\stressModifier\neq0$. This argument can be trivially modified to apply to any compact interval of time, which guarantees the existence of $\stressModifier_m>0$ if $n$ is assumed to be continuous. By evaluating the bound of \cref{eq: app: growing hoop stress} for given parameters, noting that the stress must lie above $\stressThreshold$ in order to avoid a steady state, this exponentially restricts the values of $\stressModifier_m$ needed to generate perpetually unsteady dynamics. Whilst it may be possible to construct a functional form for $\stressThreshold$ so as to satisfy these bounds and generate dynamics without a steady state, we do not expect this to occur regularly in practice due to the exponential time dependence of \cref{eq: app: growing hoop stress}. Indeed, we invariably obtain steady states as the numerical solutions to the model of \cref{sec: recovering a robust steady state} when $\springConstant=0$ across a wide range of parameter values, in support of this claim.

\end{appendices}

\bibliography{references.bib}

\begin{thebibliography}{}
\providecommand{\doi}[1]{\url{https://doi.org/#1}}
\bibcommenthead

\bibitem[\protect\citeauthoryear{Aasen, Mesnil, Naus, Lampe, and Laird}{Aasen
  et~al.}{2016}]{Aasen2016}
Aasen, T., M.~Mesnil, C.C. Naus, P.D. Lampe, and D.W. Laird. 2016, 12.
\newblock {Gap junctions and cancer: communicating for 50 years}.
\newblock {\em Nature Reviews Cancer\/}~{\em 16\/}(12): 775--788.
\newblock \doi{10.1038/nrc.2016.105} .

\bibitem[\protect\citeauthoryear{Ambrosi, Ateshian, Arruda, Cowin, Dumais,
  Goriely, Holzapfel, Humphrey, Kemkemer, Kuhl, Olberding, Taber, and
  Garikipati}{Ambrosi et~al.}{2011}]{Ambrosi2011}
Ambrosi, D., G.~Ateshian, E.~Arruda, S.~Cowin, J.~Dumais, A.~Goriely,
  G.~Holzapfel, J.~Humphrey, R.~Kemkemer, E.~Kuhl, J.~Olberding, L.~Taber, and
  K.~Garikipati. 2011, 4.
\newblock {Perspectives on biological growth and remodeling}.
\newblock {\em Journal of the Mechanics and Physics of Solids\/}~{\em 59\/}(4):
  863--883.
\newblock \doi{10.1016/j.jmps.2010.12.011} .

\bibitem[\protect\citeauthoryear{Ambrosi and Mollica}{Ambrosi and
  Mollica}{2002}]{Ambrosi2002}
Ambrosi, D. and F.~Mollica. 2002, 7.
\newblock {On the mechanics of a growing tumor}.
\newblock {\em International Journal of Engineering Science\/}~{\em 40\/}(12):
  1297--1316.
\newblock \doi{10.1016/S0020-7225(02)00014-9} .

\bibitem[\protect\citeauthoryear{Ambrosi and Mollica}{Ambrosi and
  Mollica}{2004}]{Ambrosi2004}
Ambrosi, D. and F.~Mollica. 2004, 5.
\newblock {The role of stress in the growth of a multicell spheroid}.
\newblock {\em Journal of Mathematical Biology\/}~{\em 48\/}(5): 477--499.
\newblock \doi{10.1007/s00285-003-0238-2} .

\bibitem[\protect\citeauthoryear{Ambrosi, Pezzuto, Riccobelli, Stylianopoulos,
  and Ciarletta}{Ambrosi et~al.}{2017}]{Ambrosi2017}
Ambrosi, D., S.~Pezzuto, D.~Riccobelli, T.~Stylianopoulos, and P.~Ciarletta.
  2017, 12.
\newblock {Solid Tumors Are Poroelastic Solids with a Chemo-mechanical Feedback
  on Growth}.
\newblock {\em Journal of Elasticity\/}~{\em 129\/}(1-2): 107--124.
\newblock \doi{10.1007/S10659-016-9619-9/TABLES/1} .

\bibitem[\protect\citeauthoryear{Ambrosi and Preziosi}{Ambrosi and
  Preziosi}{2009}]{Ambrosi2009}
Ambrosi, D. and L.~Preziosi. 2009, 10.
\newblock {Cell adhesion mechanisms and stress relaxation in the mechanics of
  tumours}.
\newblock {\em Biomechanics and Modeling in Mechanobiology\/}~{\em 8\/}(5):
  397--413.
\newblock \doi{10.1007/s10237-008-0145-y} .

\bibitem[\protect\citeauthoryear{Ambrosi, Preziosi, and Vitale}{Ambrosi
  et~al.}{2012}]{Ambrosi2012a}
Ambrosi, D., L.~Preziosi, and G.~Vitale. 2012, 6.
\newblock {The interplay between stress and growth in solid tumors}.
\newblock {\em Mechanics Research Communications\/}~42: 87--91.
\newblock \doi{10.1016/J.MECHRESCOM.2012.01.002} .

\bibitem[\protect\citeauthoryear{Araujo and McElwain}{Araujo and
  McElwain}{2004}]{Araujo2004}
Araujo, R.P. and D.L. McElwain. 2004, 9.
\newblock {A history of the study of solid tumour growth: The contribution of
  mathematical modelling}.
\newblock {\em Bulletin of Mathematical Biology 2004 66:5\/}~{\em 66\/}(5):
  1039--1091.
\newblock \doi{10.1016/J.BULM.2003.11.002} .

\bibitem[\protect\citeauthoryear{Bull and Byrne}{Bull and
  Byrne}{2022}]{Bull2022}
Bull, J.A. and H.M. Byrne. 2022, 5.
\newblock {The Hallmarks of Mathematical Oncology}.
\newblock {\em Proceedings of the IEEE\/}~{\em 110\/}(5): 523--540.
\newblock \doi{10.1109/JPROC.2021.3136715} .

\bibitem[\protect\citeauthoryear{Byrne}{Byrne}{2003}]{Byrne2003}
Byrne, H. 2003, 12.
\newblock {Modelling solid tumour growth using the theory of mixtures}.
\newblock {\em Mathematical Medicine and Biology\/}~{\em 20\/}(4): 341--366.
\newblock \doi{10.1093/imammb/20.4.341} .

\bibitem[\protect\citeauthoryear{Byrne and Drasdo}{Byrne and
  Drasdo}{2009}]{Byrne2009}
Byrne, H. and D.~Drasdo. 2009, 4.
\newblock {Individual-based and continuum models of growing cell populations: a
  comparison}.
\newblock {\em Journal of Mathematical Biology\/}~{\em 58\/}(4-5): 657--687.
\newblock \doi{10.1007/s00285-008-0212-0} .

\bibitem[\protect\citeauthoryear{Chen, Byrne, and King}{Chen
  et~al.}{2001}]{Chen2001}
Chen, C.Y., H.M. Byrne, and J.R. King. 2001, 9.
\newblock {The influence of growth-induced stress from the surrounding medium
  on the development of multicell spheroids}.
\newblock {\em Journal of Mathematical Biology\/}~{\em 43\/}(3): 191--220.
\newblock \doi{10.1007/s002850100091} .

\bibitem[\protect\citeauthoryear{Cheng, Tse, Jain, and Munn}{Cheng
  et~al.}{2009}]{Cheng2009}
Cheng, G., J.~Tse, R.K. Jain, and L.L. Munn. 2009, 2.
\newblock {Micro-Environmental Mechanical Stress Controls Tumor Spheroid Size
  and Morphology by Suppressing Proliferation and Inducing Apoptosis in Cancer
  Cells}.
\newblock {\em PLoS ONE\/}~{\em 4\/}(2): e4632.
\newblock \doi{10.1371/journal.pone.0004632} .

\bibitem[\protect\citeauthoryear{Ciarletta, Ambrosi, Maugin, and
  Preziosi}{Ciarletta et~al.}{2013}]{Ciarletta2013a}
Ciarletta, P., D.~Ambrosi, G.A. Maugin, and L.~Preziosi. 2013.
\newblock {Mechano-transduction in tumour growth modelling Physical constraints
  of morphogenesis and evolution. Guest editors: V. Fleury, P. Fran{\c{c}}ois
  and M-C. Ho Ba Tho}.
\newblock {\em European Physical Journal E\/}~{\em 36\/}(3).
\newblock \doi{10.1140/epje/i2013-13023-2} .

\bibitem[\protect\citeauthoryear{Delarue, Montel, Vignjevic, Prost, Joanny, and
  Cappello}{Delarue et~al.}{2014}]{Delarue2014}
Delarue, M., F.~Montel, D.~Vignjevic, J.~Prost, J.F. Joanny, and G.~Cappello.
  2014, 10.
\newblock {Compressive Stress Inhibits Proliferation in Tumor Spheroids through
  a Volume Limitation}.
\newblock {\em Biophysical Journal\/}~{\em 107\/}(8): 1821--1828.
\newblock \doi{10.1016/J.BPJ.2014.08.031} .

\bibitem[\protect\citeauthoryear{Giverso and Ciarletta}{Giverso and
  Ciarletta}{2016}]{Giverso2016}
Giverso, C. and P.~Ciarletta. 2016, 10.
\newblock {On the morphological stability of multicellular tumour spheroids
  growing in porous media}.
\newblock {\em The European Physical Journal E\/}~{\em 39\/}(10): 92.
\newblock \doi{10.1140/epje/i2016-16092-7} .

\bibitem[\protect\citeauthoryear{Goriely}{Goriely}{2017}]{Goriely2017}
Goriely, A. 2017.
\newblock {\em {The Mathematics and Mechanics of Biological Growth}}.
\newblock Interdisciplinary Applied Mathematics. Springer New York.

\bibitem[\protect\citeauthoryear{Greenspan}{Greenspan}{1972}]{Greenspan1972}
Greenspan, H.P. 1972, 12.
\newblock {Models for the Growth of a Solid Tumor by Diffusion}.
\newblock {\em Studies in Applied Mathematics\/}~{\em 51\/}(4): 317--340.
\newblock \doi{10.1002/sapm1972514317} .

\bibitem[\protect\citeauthoryear{Guillaume, Rigal, Fehrenbach, Severac,
  Ducommun, and Lobjois}{Guillaume et~al.}{2019}]{Guillaume2019}
Guillaume, L., L.~Rigal, J.~Fehrenbach, C.~Severac, B.~Ducommun, and
  V.~Lobjois. 2019, 12.
\newblock {Characterization of the physical properties of tumor-derived
  spheroids reveals critical insights for pre-clinical studies}.
\newblock {\em Scientific Reports\/}~{\em 9\/}(1): 6597.
\newblock \doi{10.1038/s41598-019-43090-0} .

\bibitem[\protect\citeauthoryear{Hanahan and Weinberg}{Hanahan and
  Weinberg}{2000}]{Hanahan2000}
Hanahan, D. and R.A. Weinberg. 2000, 1.
\newblock {The Hallmarks of Cancer}.
\newblock {\em Cell\/}~{\em 100\/}(1): 57--70.
\newblock \doi{10.1016/S0092-8674(00)81683-9} .

\bibitem[\protect\citeauthoryear{Hanahan and Weinberg}{Hanahan and
  Weinberg}{2011}]{Hanahan2011}
Hanahan, D. and R.A. Weinberg. 2011, 3.
\newblock {Hallmarks of cancer: The next generation}.
\newblock {\em Cell\/}~{\em 144\/}(5): 646--674.
\newblock \doi{10.1016/j.cell.2011.02.013} .

\bibitem[\protect\citeauthoryear{Helmlinger, Netti, Lichtenbeld, Melder, and
  Jain}{Helmlinger et~al.}{1997}]{Helmlinger1997}
Helmlinger, G., P.A. Netti, H.C. Lichtenbeld, R.J. Melder, and R.K. Jain. 1997.
\newblock {Solid stress inhibits the growth of multicellular tumor spheroids}.
\newblock {\em Nature Biotechnology\/}~{\em 15\/}(8): 778--783.
\newblock \doi{10.1038/nbt0897-778} .

\bibitem[\protect\citeauthoryear{Hirschhaeuser, Menne, Dittfeld, West,
  Mueller-Klieser, and Kunz-Schughart}{Hirschhaeuser
  et~al.}{2010}]{Hirschhaeuser2010}
Hirschhaeuser, F., H.~Menne, C.~Dittfeld, J.~West, W.~Mueller-Klieser, and L.A.
  Kunz-Schughart. 2010, 7.
\newblock {Multicellular tumor spheroids: An underestimated tool is catching up
  again}.
\newblock {\em Journal of Biotechnology\/}~{\em 148\/}(1): 3--15.
\newblock \doi{10.1016/j.jbiotec.2010.01.012} .

\bibitem[\protect\citeauthoryear{Kimpton, Walker, Hall, Bintu, Crosby, Byrne,
  and Goriely}{Kimpton et~al.}{2021}]{Kimpton2021}
Kimpton, L.S., B.J. Walker, C.L. Hall, B.~Bintu, D.~Crosby, H.M. Byrne, and
  A.~Goriely. 2021, 1.
\newblock {A Morphoelastic Shell Model of the Eye}.
\newblock {\em Journal of Elasticity\/}.
\newblock \doi{10.1007/s10659-020-09812-6} .

\bibitem[\protect\citeauthoryear{Kolosnjaj-Tabi, Gibot, Fourquaux, Golzio, and
  Rols}{Kolosnjaj-Tabi et~al.}{2019}]{Kolosnjaj-Tabi2019}
Kolosnjaj-Tabi, J., L.~Gibot, I.~Fourquaux, M.~Golzio, and M.P. Rols. 2019, 1.
\newblock {Electric field-responsive nanoparticles and electric fields:
  physical, chemical, biological mechanisms and therapeutic prospects}.
\newblock {\em Advanced Drug Delivery Reviews\/}~138: 56--67.
\newblock \doi{10.1016/j.addr.2018.10.017} .

\bibitem[\protect\citeauthoryear{Kuhl}{Kuhl}{2014}]{Kuhl2014}
Kuhl, E. 2014, 1.
\newblock {Growing matter: A review of growth in living systems}.
\newblock {\em Journal of the Mechanical Behavior of Biomedical
  Materials\/}~29: 529--543.
\newblock \doi{10.1016/j.jmbbm.2013.10.009} .

\bibitem[\protect\citeauthoryear{Kunz‐Schughart, Kreutz, and
  Knuechel}{Kunz‐Schughart et~al.}{1998}]{Kunz-Schughart1998}
Kunz‐Schughart, L.A., M.~Kreutz, and R.~Knuechel. 1998, 2.
\newblock {Multicellular spheroids: a three‐dimensional in vitro culture
  system to study tumour biology}.
\newblock {\em International Journal of Experimental Pathology\/}~{\em
  79\/}(1): 1--23.
\newblock \doi{10.1046/j.1365-2613.1998.00051.x} .

\bibitem[\protect\citeauthoryear{Maia, Caja, Strano~Moraes, Couto, and
  Costa-Silva}{Maia et~al.}{2018}]{Maia2018}
Maia, J., S.~Caja, M.C. Strano~Moraes, N.~Couto, and B.~Costa-Silva. 2018, 2.
\newblock {Exosome-Based Cell-Cell Communication in the Tumor
  Microenvironment}.
\newblock {\em Frontiers in Cell and Developmental Biology\/}~{\em 6\/}(FEB):
  1--19.
\newblock \doi{10.3389/fcell.2018.00018} .

\bibitem[\protect\citeauthoryear{Murphy, Browning, Gunasingh, Haass, and
  Simpson}{Murphy et~al.}{2022}]{Murphy2022}
Murphy, R.J., A.P. Browning, G.~Gunasingh, N.K. Haass, and M.J. Simpson. 2022,
  12.
\newblock {Designing and interpreting 4D tumour spheroid experiments}.
\newblock {\em Communications Biology\/}~{\em 5\/}(1): 91.
\newblock \doi{10.1038/s42003-022-03018-3} .

\bibitem[\protect\citeauthoryear{Nia, Datta, Seano, Huang, Munn, and Jain}{Nia
  et~al.}{2018}]{Nia2018}
Nia, H.T., M.~Datta, G.~Seano, P.~Huang, L.L. Munn, and R.K. Jain. 2018, 5.
\newblock {Quantifying solid stress and elastic energy from excised or in situ
  tumors}.
\newblock {\em Nature Protocols\/}~{\em 13\/}(5): 1091--1105.
\newblock \doi{10.1038/nprot.2018.020} .

\bibitem[\protect\citeauthoryear{Northcott, Dean, Mouw, and Weaver}{Northcott
  et~al.}{2018}]{Northcott2018}
Northcott, J.M., I.S. Dean, J.K. Mouw, and V.M. Weaver. 2018, 2.
\newblock {Feeling stress: The mechanics of cancer progression and aggression}.
\newblock {\em Frontiers in Cell and Developmental Biology\/}~{\em 6\/}(FEB):
  17.
\newblock \doi{10.3389/fcell.2018.00017} .

\bibitem[\protect\citeauthoryear{Pavlova and Thompson}{Pavlova and
  Thompson}{2016}]{Pavlova2016}
Pavlova, N.N. and C.B. Thompson. 2016, 1.
\newblock {The Emerging Hallmarks of Cancer Metabolism}.
\newblock {\em Cell Metabolism\/}~{\em 23\/}(1): 27--47.
\newblock \doi{10.1016/j.cmet.2015.12.006} .

\bibitem[\protect\citeauthoryear{Rodriguez, Hoger, and McCulloch}{Rodriguez
  et~al.}{1994}]{Rodriguez1994}
Rodriguez, E.K., A.~Hoger, and A.D. McCulloch. 1994, 4.
\newblock {Stress-dependent finite growth in soft elastic tissues}.
\newblock {\em Journal of Biomechanics\/}~{\em 27\/}(4): 455--467.
\newblock \doi{10.1016/0021-9290(94)90021-3} .

\bibitem[\protect\citeauthoryear{Roose, Chapman, and Maini}{Roose
  et~al.}{2007}]{Roose2007}
Roose, T., S.J. Chapman, and P.K. Maini. 2007, 1.
\newblock {Mathematical Models of Avascular Tumor Growth}.
\newblock {\em SIAM Review\/}~{\em 49\/}(2): 179--208.
\newblock \doi{10.1137/S0036144504446291} .

\bibitem[\protect\citeauthoryear{Roose, Netti, Munn, Boucher, and Jain}{Roose
  et~al.}{2003}]{Roose2003}
Roose, T., P.A. Netti, L.L. Munn, Y.~Boucher, and R.K. Jain. 2003, 11.
\newblock {Solid stress generated by spheroid growth estimated using a linear
  poroelasticity model}.
\newblock {\em Microvascular Research\/}~{\em 66\/}(3): 204--212.
\newblock \doi{10.1016/S0026-2862(03)00057-8} .

\bibitem[\protect\citeauthoryear{Sengupta and Balla}{Sengupta and
  Balla}{2018}]{Sengupta2018}
Sengupta, S. and V.K. Balla. 2018, 11.
\newblock {A review on the use of magnetic fields and ultrasound for
  non-invasive cancer treatment}.
\newblock {\em Journal of Advanced Research\/}~14: 97--111.
\newblock \doi{10.1016/j.jare.2018.06.003} .

\bibitem[\protect\citeauthoryear{Sherratt and Chaplain}{Sherratt and
  Chaplain}{2001}]{Sherratt2001}
Sherratt, J.A. and M.A. Chaplain. 2001, 10.
\newblock {A new mathematical model for avascular tumour growth}.
\newblock {\em Journal of Mathematical Biology\/}~{\em 43\/}(4): 291--312.
\newblock \doi{10.1007/s002850100088} .

\bibitem[\protect\citeauthoryear{Stylianopoulos, Martin, Chauhan, Jain,
  Diop-Frimpong, Bardeesy, Smith, Ferrone, Hornicek, Boucher, Munn, and
  Jain}{Stylianopoulos et~al.}{2012}]{Stylianopoulos2012}
Stylianopoulos, T., J.D. Martin, V.P. Chauhan, S.R. Jain, B.~Diop-Frimpong,
  N.~Bardeesy, B.L. Smith, C.R. Ferrone, F.J. Hornicek, Y.~Boucher, L.L. Munn,
  and R.K. Jain. 2012, 9.
\newblock {Causes, consequences, and remedies for growth-induced solid stress
  in murine and human tumors}.
\newblock {\em Proceedings of the National Academy of Sciences\/}~{\em
  109\/}(38): 15101--15108.
\newblock \doi{10.1073/pnas.1213353109} .

\bibitem[\protect\citeauthoryear{Sung, Ferlay, Siegel, Laversanne,
  Soerjomataram, Jemal, and Bray}{Sung et~al.}{2021}]{Sung2021}
Sung, H., J.~Ferlay, R.L. Siegel, M.~Laversanne, I.~Soerjomataram, A.~Jemal,
  and F.~Bray. 2021, 5.
\newblock {Global Cancer Statistics 2020: GLOBOCAN Estimates of Incidence and
  Mortality Worldwide for 36 Cancers in 185 Countries}.
\newblock {\em CA: A Cancer Journal for Clinicians\/}~{\em 71\/}(3): 209--249.
\newblock \doi{10.3322/caac.21660} .

\bibitem[\protect\citeauthoryear{Vaupel, Kallinowski, and Okunieff}{Vaupel
  et~al.}{1989}]{Vaupel1989}
Vaupel, P., F.~Kallinowski, and P.~Okunieff. 1989.
\newblock {Blood flow, oxygen and nutrient supply, and metabolic
  microenvironment of human tumors: a review}.
\newblock {\em Cancer research\/}~{\em 49\/}(23): 6449--6465 .

\bibitem[\protect\citeauthoryear{Ward and King}{Ward and King}{1997}]{Ward1997}
Ward, J.P. and J.R. King. 1997.
\newblock {Mathematical modelling of avascular-tumour growth}.
\newblock {\em Mathematical Medicine and Biology: A Journal of the IMA\/}~{\em
  14\/}(1): 39--69 .

\bibitem[\protect\citeauthoryear{Yan, Ramirez-Guerrero, Lowengrub, and Wu}{Yan
  et~al.}{2021}]{Yan2021}
Yan, H., D.~Ramirez-Guerrero, J.~Lowengrub, and M.~Wu. 2021, 12.
\newblock {Stress generation, relaxation and size control in confined tumor
  growth}.
\newblock {\em PLOS Computational Biology\/}~{\em 17\/}(12): e1009701.
\newblock \doi{10.1371/journal.pcbi.1009701} .

\end{thebibliography}

\end{document}